
%




\font\bigbf=cmbx10 scaled\magstep2

\font\twelverm=cmr10 scaled 1200    \font\twelvei=cmmi10 scaled 1200
\font\twelvesy=cmsy10 scaled 1200   \font\twelveex=cmex10 scaled 1200
\font\twelvebf=cmbx10 scaled 1200   \font\twelvesl=cmsl10 scaled 1200
\font\twelvett=cmtt10 scaled 1200   \font\twelveit=cmti10 scaled 1200

\skewchar\twelvei='177   \skewchar\twelvesy='60


\def\twelvepoint{\normalbaselineskip=12.4pt
  \abovedisplayskip 12.4pt plus 3pt minus 9pt
  \belowdisplayskip 12.4pt plus 3pt minus 9pt
  \abovedisplayshortskip 0pt plus 3pt
  \belowdisplayshortskip 7.2pt plus 3pt minus 4pt
  \smallskipamount=3.6pt plus1.2pt minus1.2pt
  \medskipamount=7.2pt plus2.4pt minus2.4pt
  \bigskipamount=14.4pt plus4.8pt minus4.8pt
  \def\rm{\fam0\twelverm}          \def\it{\fam\itfam\twelveit}%
  \def\sl{\fam\slfam\twelvesl}     \def\bf{\fam\bffam\twelvebf}%
  \def\mit{\fam 1}                 \def\cal{\fam 2}%
  \def\tt{\twelvett}
  \textfont0=\twelverm   \scriptfont0=\tenrm   \scriptscriptfont0=\sevenrm
  \textfont1=\twelvei    \scriptfont1=\teni    \scriptscriptfont1=\seveni
  \textfont2=\twelvesy   \scriptfont2=\tensy   \scriptscriptfont2=\sevensy
  \textfont3=\twelveex   \scriptfont3=\twelveex
 \scriptscriptfont3=\twelveex
  \textfont\itfam=\twelveit
  \textfont\slfam=\twelvesl
  \textfont\bffam=\twelvebf \scriptfont\bffam=\tenbf
  \scriptscriptfont\bffam=\sevenbf
  \normalbaselines\rm}



\def\beginlinemode{\endmode
  \begingroup\parskip=0pt
\obeylines\def\\{\par}\def\endmode{\par\endgroup}}
\def\beginparmode{\endmode
  \begingroup \def\endmode{\par\endgroup}}
\let\endmode=\par
{\obeylines\gdef\
{}}
\def\singlespace{\baselineskip=\normalbaselineskip}
\def\oneandathirdspace{\baselineskip=\normalbaselineskip
  \multiply\baselineskip by 4 \divide\baselineskip by 3}

\def\doublespace{\baselineskip=
\normalbaselineskip \multiply\baselineskip by 2}

\newcount\firstpageno
\firstpageno=1
\footline={\ifnum\pageno<\firstpageno{\hfil}%
\else{\hfil\twelverm\folio\hfil}\fi}
\let\rawfootnote=\footnote              
\def\footnote#1#2{{\rm\singlespace\parindent=0pt\rawfootnote{#1}{#2}}}
\def\raggedcenter{\leftskip=4em plus 12em \rightskip=\leftskip
  \parindent=0pt \parfillskip=0pt \spaceskip=.3333em \xspaceskip=.5em
  \pretolerance=9999 \tolerance=9999
  \hyphenpenalty=9999 \exhyphenpenalty=9999 }
\def\dateline{\rightline{\ifcase\month\or
  January\or February\or March\or April\or May\or June\or
  July\or August\or September\or October\or November\or December\fi
  \space\number\year}}
\def\received{\vskip 3pt plus 0.2fill
 \centerline{\sl (Received\space\ifcase\month\or
  January\or February\or March\or April\or May\or June\or
  July\or August\or September\or October\or November\or December\fi
  \qquad, \number\year)}}


\hsize=6.5truein
\vsize=8.9truein
\voffset=0.0truein
\parskip=\medskipamount
\twelvepoint            
\oneandathirdspace           
\overfullrule=0pt       



\def\title                      
  {\null\vskip 3pt plus 0.2fill
   \beginlinemode \doublespace \raggedcenter \bigbf}

\def\author                     
  {\vskip 3pt plus 0.2fill \beginlinemode
   \singlespace \raggedcenter}

\def\affil                      
  {\vskip 4pt 
\beginlinemode
   \singlespace \raggedcenter \sl}

\def\abstract                   
  {\vskip 3pt plus 0.3fill \beginparmode
   \oneandathirdspace\narrower}

\def\endtitlepage               
  {\endpage                     
   \body}

\def\body                       
  {\beginparmode}               

\def\subhead#1{                 
  \vskip 0.25truein             
  \noindent{{\it {#1}} \par}
   \nobreak\vskip 0.15truein\nobreak}

\def\refto#1{[#1]}           

\def\references                 
  {\subhead{\bf References}         
   \beginparmode
   \frenchspacing \parindent=0pt \leftskip=1truecm
   \oneandathirdspace\parskip=8pt plus 3pt
 \everypar{\hangindent=\parindent}}

\gdef\refis#1{\indent\hbox to 0pt{\hss#1.~}}    

\gdef\journal#1, #2, #3, 1#4#5#6{               
    {\sl #1~}{\bf #2}, #3 (1#4#5#6)}           

\def\refstylenp{                
  \gdef\refto##1{ [##1]}                                
  \gdef\refis##1{\indent\hbox to 0pt{\hss##1)~}}        
  \gdef\journal##1, ##2, ##3, ##4 {                     
     {\sl ##1~}{\bf ##2~}(##3) ##4 }}

\def\refstyleprnp{              
  \gdef\refto##1{ [##1]}                                
  \gdef\refis##1{\indent\hbox to 0pt{\hss##1)~}}        
  \gdef\journal##1, ##2, ##3, 1##4##5##6{               
    {\sl ##1~}{\bf ##2~}(1##4##5##6) ##3}}

\def\prd{\journal Phys. Rev. D, }

\def\prl{\journal Phys. Rev. Lett., }

\def\jmp{\journal J. Math. Phys., }

\def\cqg{\journal Class. Quantum Grav., }

\def\ann{\journal Ann. Phys., }

\def\endreferences{\body}

\def\figurecaptions             
  { \beginparmode
   \subhead{Figure Captions}
}

\def\endpage                    
  {\vfill\eject}

\def\endpaper                   
  {\endmode\vfill\supereject}

\def\hook{\mathbin{\raise2.5pt\hbox{\hbox{{\vbox{\hrule height.4pt
width6pt depth0pt}}}\vrule height3pt width.4pt depth0pt}\,}}
\def\ref#1{Ref. #1}                     
\def\Ref#1{Ref. #1}                     

\def\frac#1#2{{\textstyle{#1 \over #2}}}

\def\ie{{\it i.e.,\ }}

\def\etc{{\it etc.}}

\def\sla{\raise.15ex\hbox{$/$}\kern-.57em}
\def\leaderfill{\leaders\hbox to 1em{\hss.\hss}\hfill}
\def\twiddle{\lower.9ex\rlap{$\kern-.1em\scriptstyle\sim$}}
\def\bigtwiddle{\lower1.ex\rlap{$\sim$}}
\def\gtwid{
\mathrel{\raise.3ex\hbox{$>$\kern-.75em\lower1ex\hbox{$\sim$}}}}
\def\ltwid{\mathrel{\raise.3ex\hbox
{$<$\kern-.75em\lower1ex\hbox{$\sim$}}}}
\def\square{\kern1pt\vbox{\hrule height 1.2pt\hbox
{\vrule width 1.2pt\hskip 3pt
   \vbox{\vskip 6pt}\hskip 3pt\vrule width 0.6pt}
\hrule height 0.6pt}\kern1pt}

\def\m@th{\mathsurround=0pt }
\def\leftrightarrowfill{$\m@th \mathord\leftarrow \mkern-6mu
 \cleaders\hbox{$\mkern-2mu \mathord- \mkern-2mu$}\hfill
 \mkern-6mu \mathord\rightarrow$}
\def\overleftrightarrow#1{\vbox{\ialign{##\crcr
     \leftrightarrowfill\crcr\noalign{\kern-1pt\nointerlineskip}
     $\hfil\displaystyle{#1}\hfil$\crcr}}}


\font\titlefont=cmr10 scaled\magstep3

\def\martinstyletitle                      
  {\null\vskip 3pt plus 0.2fill
   \beginlinemode \doublespace \raggedcenter \titlefont}

\font\twelvesc=cmcsc10 scaled 1200

\def\author                     
  {\vskip 3pt plus 0.2fill \beginlinemode
   \singlespace \raggedcenter\twelvesc}


\def\heading                            
  {\vskip 0.5truein plus 0.1truein      
\endheading
   \beginparmode \def\\{\par} \parskip=0pt \singlespace \raggedcenter}

\def\endheading
  {\par\nobreak\vskip 0.25truein\nobreak\beginparmode}

\def\subheading                         
  {\vskip 0.25truein plus 0.1truein
   \beginlinemode \singlespace \parskip=0pt \def\\{\par}\raggedcenter}

\def\tag#1$${\eqno(#1)$$}

\def\align#1$${\eqalign{#1}$$}

\def\aligntag#1$${\gdef\tag##1\\{&(##1)\cr}\eqalignno{#1\\}$$
  \gdef\tag##1$${\eqno(##1)$$}}

\def\endaligntag{}

\def\overset #1\to#2{{\mathop{#2}\limits^{#1}}}
\def\underset#1\to#2{{\let\next=#1\mathpalette\undersetpalette#2}}
\def\undersetpalette#1#2{\vtop{\baselineskip0pt
\ialign{$\mathsurround=0pt #1\hfil##\hfil$\crcr#2\crcr\next\crcr}}}


\def\ref#1{Ref.~#1}                     
\def\Ref#1{Ref.~#1}                     
\def\[#1]{[\cite{#1}]}
\def\cite#1{{#1}}
\def\(#1){(\call{#1})}
\def\call#1{{#1}}
\def\taghead#1{}
\def\frac#1#2{{#1 \over #2}}

\def\12{{1\over2}}

\def\ie{{\it i.e.,\ }}

\def\etc{{\it etc.\ }}

\def\sla{\raise.15ex\hbox{$/$}\kern-.57em}
\def\leaderfill{\leaders\hbox to 1em{\hss.\hss}\hfill}
\def\twiddle{\lower.9ex\rlap{$\kern-.1em\scriptstyle\sim$}}
\def\bigtwiddle{\lower1.ex\rlap{$\sim$}}
\def\gtwid{\mathrel{\raise.3ex\hbox{$>$
\kern-.75em\lower1ex\hbox{$\sim$}}}}
\def\ltwid{\mathrel{\raise.3ex\hbox{$<$
\kern-.75em\lower1ex\hbox{$\sim$}}}}
\def\square{\kern1pt\vbox{\hrule height 1.2pt\hbox
{\vrule width 1.2pt\hskip 3pt
   \vbox{\vskip 6pt}\hskip 3pt\vrule width 0.6pt}
\hrule height 0.6pt}\kern1pt}
\def\tdot#1{\mathord{\mathop{#1}\limits^{\kern2pt\ldots}}}

\def\pmb#1{\setbox0=\hbox{#1}%
  \kern-.025em\copy0\kern-\wd0
  \kern  .05em\copy0\kern-\wd0
  \kern-.025em\raise.0433em\box0 }

\catcode`@=11
\newcount\tagnumber\tagnumber=0

\immediate\newwrite\eqnfile
\newif\if@qnfile\@qnfilefalse
\def\write@qn#1{}
\def\writenew@qn#1{}
\def\w@rnwrite#1{\write@qn{#1}\message{#1}}
\def\@rrwrite#1{\write@qn{#1}\errmessage{#1}}

\def\taghead#1{\gdef\t@ghead{#1}\global\tagnumber=0}
\def\t@ghead{}

\expandafter\def\csname @qnnum-3\endcsname
  {{\t@ghead\advance\tagnumber by -3\relax\number\tagnumber}}
\expandafter\def\csname @qnnum-2\endcsname
  {{\t@ghead\advance\tagnumber by -2\relax\number\tagnumber}}
\expandafter\def\csname @qnnum-1\endcsname
  {{\t@ghead\advance\tagnumber by -1\relax\number\tagnumber}}
\expandafter\def\csname @qnnum0\endcsname
  {\t@ghead\number\tagnumber}
\expandafter\def\csname @qnnum+1\endcsname
  {{\t@ghead\advance\tagnumber by 1\relax\number\tagnumber}}
\expandafter\def\csname @qnnum+2\endcsname
  {{\t@ghead\advance\tagnumber by 2\relax\number\tagnumber}}
\expandafter\def\csname @qnnum+3\endcsname
  {{\t@ghead\advance\tagnumber by 3\relax\number\tagnumber}}

\def\equationfile{%
  \@qnfiletrue\immediate\openout\eqnfile=\jobname.eqn%
  \def\write@qn##1{\if@qnfile\immediate\write\eqnfile{##1}\fi}
  \def\writenew@qn##1{\if@qnfile\immediate\write\eqnfile
    {\noexpand\tag{##1} = (\t@ghead\number\tagnumber)}\fi}
}

\def\callall#1{\xdef#1##1{#1{\noexpand\call{##1}}}}
\def\call#1{\each@rg\callr@nge{#1}}

\def\each@rg#1#2{{\let\thecsname=#1\expandafter\first@rg#2,\end,}}
\def\first@rg#1,{\thecsname{#1}\apply@rg}
\def\apply@rg#1,{\ifx\end#1\let\next=\relax%
\else,\thecsname{#1}\let\next=\apply@rg\fi\next}

\def\callr@nge#1{\calldor@nge#1-\end-}
\def\callr@ngeat#1\end-{#1}
\def\calldor@nge#1-#2-{\ifx\end#2\@qneatspace#1 %
  \else\calll@@p{#1}{#2}\callr@ngeat\fi}
\def\calll@@p#1#2{\ifnum#1>#2{\@rrwrite
{Equation range #1-#2\space is bad.}
\errhelp{If you call a series of equations by the notation M-N, then M and
N must be integers, and N must be greater than or equal to M.}}\else %
{\count0=#1\count1=
#2\advance\count1 by1\relax\expandafter\@qncall\the\count0,%
  \loop\advance\count0 by1\relax%
    \ifnum\count0<\count1,\expandafter\@qncall\the\count0,%
  \repeat}\fi}

\def\@qneatspace#1#2 {\@qncall#1#2,}
\def\@qncall#1,{\ifunc@lled{#1}{\def\next{#1}\ifx\next\empty\else
  \w@rnwrite{Equation number \noexpand\(>>#1<<)
has not been defined yet.}
  >>#1<<\fi}\else\csname @qnnum#1\endcsname\fi}

\let\eqnono=\eqno
\def\eqno(#1){\tag#1}
\def\tag#1$${\eqnono(\displayt@g#1 )$$}

\def\aligntag#1\endaligntag
  $${\gdef\tag##1\\{&(##1 )\cr}\eqalignno{#1\\}$$
  \gdef\tag##1$${\eqnono(\displayt@g##1 )$$}}

\def\eqalignno#1{\displ@y \tabskip\centering
  \halign to\displaywidth{\hfil$\displaystyle{##}$\tabskip\z@skip
    &$\displaystyle{{}##}$\hfil\tabskip\centering
    &\llap{$\displayt@gpar##$}\tabskip\z@skip\crcr
    #1\crcr}}

\def\displayt@gpar(#1){(\displayt@g#1 )}

\def\displayt@g#1 {\rm\ifunc@lled{#1}\global\advance\tagnumber by1
        {\def\next{#1}\ifx\next\empty\else\expandafter
        \xdef\csname
 @qnnum#1\endcsname{\t@ghead\number\tagnumber}\fi}%
  \writenew@qn{#1}\t@ghead\number\tagnumber\else
        {\edef\next{\t@ghead\number\tagnumber}%
        \expandafter\ifx\csname @qnnum#1\endcsname\next\else
        \w@rnwrite{Equation \noexpand\tag{#1} is
a duplicate number.}\fi}%
  \csname @qnnum#1\endcsname\fi}

\def\ifunc@lled#1{\expandafter\ifx\csname @qnnum#1\endcsname\relax}

\let\@qnend=\end\gdef\end{\if@qnfile
\immediate\write16{Equation numbers
written on []\jobname.EQN.}\fi\@qnend}

\catcode`@=12

\catcode`@=11
\newcount\r@fcount \r@fcount=0
\newcount\r@fcurr
\immediate\newwrite\reffile
\newif\ifr@ffile\r@ffilefalse
\def\w@rnwrite#1{\ifr@ffile\immediate\write\reffile{#1}\fi\message{#1}}

\def\writer@f#1>>{}
\def\referencefile{
  \r@ffiletrue\immediate\openout\reffile=\jobname.ref%
  \def\writer@f##1>>{\ifr@ffile\immediate\write\reffile%
    {\noexpand\refis{##1} = \csname r@fnum##1\endcsname = %
     \expandafter\expandafter\expandafter\strip@t\expandafter%
     \meaning\csname r@ftext
\csname r@fnum##1\endcsname\endcsname}\fi}%
  \def\strip@t##1>>{}}

\def\citeall#1{\xdef#1##1{#1{\noexpand\cite{##1}}}}
\def\cite#1{\each@rg\citer@nge{#1}}	

\def\each@rg#1#2{{\let\thecsname=#1\expandafter\first@rg#2,\end,}}
\def\first@rg#1,{\thecsname{#1}\apply@rg}	
\def\apply@rg#1,{\ifx\end#1\let\next=\relax
\else,\thecsname{#1}\let\next=\apply@rg\fi\next}

\def\citer@nge#1{\citedor@nge#1-\end-}	
\def\citer@ngeat#1\end-{#1}
\def\citedor@nge#1-#2-{\ifx\end#2\r@featspace#1 
  \else\citel@@p{#1}{#2}\citer@ngeat\fi}	
\def\citel@@p#1#2{\ifnum#1>#2{\errmessage{Reference range #1-
#2\space is bad.}%
    \errhelp{If you cite a series of references by the notation M-N, then M
and
    N must be integers, and N must be greater than or equal to M.}}\else%
 {\count0=#1\count1=#2\advance\count1
by1\relax\expandafter\r@fcite\the\count0,
  \loop\advance\count0 by1\relax
    \ifnum\count0<\count1,\expandafter\r@fcite\the\count0,%
  \repeat}\fi}

\def\r@featspace#1#2 {\r@fcite#1#2,}	
\def\r@fcite#1,{\ifuncit@d{#1}
    \newr@f{#1}%
    \expandafter\gdef\csname r@ftext\number\r@fcount\endcsname%
                     {\message{Reference #1 to be supplied.}%
                      \writer@f#1>>#1 to be supplied.\par}%
 \fi%
 \csname r@fnum#1\endcsname}
\def\ifuncit@d#1{\expandafter\ifx\csname r@fnum#1\endcsname\relax}%
\def\newr@f#1{\global\advance\r@fcount by1%
    \expandafter\xdef\csname r@fnum#1\endcsname{\number\r@fcount}}

\let\r@fis=\refis			
\def\refis#1#2#3\par{\ifuncit@d{#1}
   \newr@f{#1}%
   \w@rnwrite{Reference #1=\number\r@fcount\space is not cited up to
 now.}\fi%
  \expandafter
\gdef\csname r@ftext\csname r@fnum#1\endcsname\endcsname%
  {\writer@f#1>>#2#3\par}}

\def\ignoreuncited{
   \def\refis##1##2##3\par{\ifuncit@d{##1}%
    \else\expandafter\gdef
\csname r@ftext\csname r@fnum##1\endcsname\endcsname%
     {\writer@f##1>>##2##3\par}\fi}}

\def\r@ferr{\endreferences\errmessage{I was expecting to see
\noexpand\endreferences before now;  I have inserted it here.}}
\let\r@ferences=\references
\def\references{\r@ferences\def\endmode{\r@ferr\par\endgroup}}

\let\endr@ferences=\endreferences
\def\endreferences{\r@fcurr=0
  {\loop\ifnum\r@fcurr<\r@fcount
    \advance\r@fcurr by
1\relax\expandafter\r@fis\expandafter{\number\r@fcurr}%
    \csname r@ftext\number\r@fcurr\endcsname%
  \repeat}\gdef\r@ferr{}\endr@ferences}


\let\r@fend=\endpaper\gdef\endpaper{\ifr@ffile
\immediate\write16{Cross References written on
[]\jobname.REF.}\fi\r@fend}

\catcode`@=12

\citeall\refto		
\citeall\ref		%
\citeall\Ref		%

\ignoreuncited
\def\psitilde{{\widetilde\psi}}
\def\ss{\scriptscriptstyle}

\def\det{{\rm det}}
\def\psit{{\tilde\psi}}
\def\phit{{\tilde\phi}}

\def\ord{{\cal O}}%
\pageno0
\line{\hfill October 1995}
\title INTERNAL TIME FORMALISM FOR SPACETIMES WITH TWO KILLING VECTORS
\author Joseph D.~Romano
\affil Department of Physics
University of Utah
Salt Lake City, UT 84112 USA
\affil Department of Physics
University of Wisconsin
Milwaukee, WI 53201 USA${}^*$
\author Charles G.~Torre
\affil Department of Physics
Utah State University
Logan, UT 84322-4415 USA
\abstract
The Hamiltonian structure of spacetimes with two commuting Killing vector
fields
is analyzed for the purpose of addressing the various problems of time that
arise in
canonical gravity.  Two specific models are considered: (i) cylindrically
symmetric spacetimes, and (ii) toroidally symmetric spacetimes, which
respectively involve open and closed universe boundary conditions.  For each
model canonical variables which can be used to identify points of space and
instants of time,
{\it i.e.}, internally defined spacetime coordinates, are identified.  To do
this it is necessary to extend the usual ADM phase space by a finite number of
degrees of freedom. Canonical transformations are exhibited that identify each
of these models with harmonic maps in the parametrized field theory formalism.
The identifications made between the gravitational models and harmonic map
field
theories are completely gauge invariant, that is, no coordinate conditions are
needed.  The degree to which the problems of time are resolved in these models
is discussed.
\bigskip\bigskip\noindent
${}^*$Current address.
\endtitlepage
%
%
\taghead{1.}
\line{\bf 1. Introduction.\hfill}

Throughout a large part of the history of general relativity much effort has
been expended toward disentangling the true degrees of freedom of the
gravitational field from the ``pure gauge'' degrees of freedom brought into the
theory via the principle of general covariance.  In the Hamiltonian form of the
theory this problem involves understanding the solution space of the initial
value constraints and the appropriate free data for the Cauchy problem.
Classically, a characterization of the true degrees of freedom is relevant for
analyzing dynamical evolution of strongly gravitating systems, {\it e.g.},
binary black hole systems, as well as for understanding fundamental issues
in relativity, {\it e.g.}, cosmic censorship. The intertwining of gauge
degrees of freedom and dynamical degrees of freedom is especially vexing in
quantum gravity, where it leads to many of the
``problems of time'' \refto{Kuchar1992}.  Evidently,
canonical quantization of the gravitational field would be expediated by a
sufficiently explicit characterization of the true degrees of freedom.  For the
most part, the
strategy for doing this stems from the original Hamiltonian description of
gravitation provided by Arnowitt, Deser, Misner \refto{ADM1962} and Dirac
\refto{Dirac1958}.  The
philosophy adopted there is that the Einstein field equations define an
``already parametrized field theory'' in which certain non-dynamical canonical
variables represent points of space and instants of time, relative to which the
true degrees of freedom evolve.
In the ADM approach, the dynamical content of relativity is exposed by
coordinate
conditions which fix the non-dynamical gauge variables.  This leads to a
desciption of gravitational dynamics relative to a fixed foliation of
spacetime,
that is, relative to a fixed family of observers.  A prominent example of this
approach appears in the conformal approach to the initial value problem
\refto{York1980}.
 From such work it seems that the ADM approach is quite adequate for
addressing many problems
in classical relativity.
However, the price paid for obtaining technical control over the true degrees
of freedom
is that general covariance is lost in the sense that one is obliged to view
dynamics from the point of view of a given set of observers.   In quantum
theory
this provides a first instance of a problem of time, which might be called
the problem of general covariance:  How to give the state of the
gravitational field on an
arbitrary hypersurface, that is, with respect to arbitrary observers?  In the
ADM approach one is prohibited from even asking this question.

An alternative approach to describing the true degrees of freedom of the
gravitational field that preserves general covariance is available.  This
approach still relies upon the possibility of extracting
``many-fingered time'' degrees of freedom---or more precisely,
spacelike embeddings of Cauchy surfaces from the gravitational phase
space, but does
not
fix the foliation with coordinate conditions.  Instead,
 one describes evolution
of the true degrees of freedom relative to an arbitrary foliation, {\it i.e.},
one casts the Einstein equations in the form of a parametrized field
theory \refto{Dirac1964}. This point of view was developed in considerable
detail by Kucha\v r
\refto{Kuchar1972}, who called it the ``bubble time'' dynamics of the
gravitational field.
Many of the problems of time are mitigated using this approach
\refto{Kuchar1992}, which is now
known as the ``internal time formalism''.
Implementation
 of the internal time formalism hinges upon the possibility of (i)
finding a canonical transformation on the gravitational phase space which
separates four canonical variables to play the role of spacelike embeddings of
a Cauchy surface into spacetime; (ii) eliminating the momenta conjugate to the
embeddings by the
initial value constraints. If this can be done, the remaining variables
represent the true degrees of freedom, whose dynamical evolution occurs via
correlation with the arbitrary spacelike slices provided by the embedding
variables.  So far, the internal time approach to gravitational dynamics has
been
implemented in lower-dimensional models, typically symmetry reductions of the
full theory.  There are a plethora of homogeneous cosmological models in which
it is possible, at least locally,
 to isolate a canonical variable representing time and then solve the
constratints for the momentum conjugate to time.  These models
possess a finite number of degrees of freedom.  Relatively few field theoretic
models, {\it i.e.}, models possessing an infinity of degrees of freedom, exist
in which one can implement the internal time program.  Such models are, to our
knowledge, always 2-dimensional field theories, most notably
Einstein-Rosen waves \refto{Kuchar1971}, plane-gravitational waves
\refto{Neville1993}, spherically symmetric vacuum
gravity \refto{Kuchar1994}, the bosonic string \refto{CGT1989,CGT1991b}, and
related models \refto{CGT1989a}.  For the most part,
these models are generally covariant theories of one or more free fields,
{\it e.g.}, the dynamics of the Einstein-Rosen wave is that of a single free
scalar field representing the metric amplitude of the wave.

Our goal in this paper is to establish an internal time formalism for the
reduction of the vacuum
Einstein equations obtained by assuming the existence of two commuting
spacelike Killing vector fields.  We consider models involving both open
universe and
closed universe boundary conditions. Our work generalizes previous models, in
particular the Einstein-Rosen waves
\refto{Kuchar1971}, in two key ways.
First, the reduced system of equations describes a pair of {\it interacting}
fields.  To our knowledge, this is the only symmetry reduction of the vacuum
Einstein equations known that admits an internal time formulation and in which
the true
degrees of freedom constitute a non-linear field theory.  Second, in terms of
the two Killing vector model, we illustrate in detail the fact that, strictly
speaking, general relativity is not an already parametrized field theory.  In
order to extract canonical variables representing embeddings from the ADM phase
space it is necessary to extend that phase space by a finite number of degrees
of freedom.  For open universes, the additional non-gravitational variables
represent the asymptotic location of a spacelike hypersurface relative to an
inertial frame ``at infinity'' \refto{Regge1974}.  For closed universes,
 the intrinsic and extrinsic geometry of a
hypersurface are inadequate to define the embedding of that hypersurface in
spacetime \refto{CGT1992a}; additional non-gravitational variables are needed
to accomplish
this.  Given the extensions of the phase space, the resulting dynamical system
is in each case a generally covariant (parametrized) formulation of harmonic
maps from a flat 3-dimensional spacetime to a 2-dimensional target space of
constant negative curvature.

In the next section we define the models.  Our open universe model is obtained
by imposing cylindrical symmetry on spacetime.  Our closed universe model,
first
considered by Gowdy \refto{Gowdy1974}, defines the universe to
be a 3-torus which admits a 2-torus group of isometries.  We do not impose
hypersurface orthogonality (reflection symmetry) on the Killing vectors and the
resulting dynamical theory is in each case
 intrinsically non-linear, in contrast to the
Einstein-Rosen waves \refto{Kuchar1971} or ``polarized'' Gowdy models
\refto{Gowdy1971}.  In the language of the
linearized theory, both gravitational polarizations propagate and interact.
In
\S3 we exhibit canonical transformations, inspired by \refto{Kuchar1971} and
\refto{CGT1989}, which allow
us to extract
the embeddings, solve the constraints, and reveal the true degrees of freedom
in each model.  Other two Killing vector models, which differ in the choice of
isometry group, spacetime topology, {\etc}, can be treated in much the same way
as we do here.  As mentioned above, these results lead to substantial
simplifications in the problems of time that occur in canonical quantum
gravity.
We discuss these problems and the general structure of
the quantum theory based on
the internal time formulation of the two Killing vector models in \S4.  In the
cylindrically symmetric model it is necessary to keep track of the behavior of
fields on the
axis
of symmetry and at infinity.  We summarize our boundary and fall-off conditions
in an appendix, which is referred to throughout the paper.
\bigskip
\taghead{2.} \line{\bf 2. Spacetimes with two commuting Killing
vectors.\hfill}

We are going to study the Hamiltonian structure of spacetimes $({\cal
M},g)$ which admit two linearly independent spacelike Killing vector fields
$K_A^\alpha$,
$A=1,2$.
The Killing vector fields are assumed to commute:
$$
 [K_A,K_B]^\alpha=0,\tag 21
$$ and to generate a two-dimensional ``orthogonally transitive''
group $\cal G$ of isometries.  This latter requirement means that the
distribution of two dimensional vector spaces spanned by $K_A^\alpha$ at each
point is orthogonal to a foliation of $\cal M$ by surfaces
$M\hookrightarrow{\cal M}$.   Our assumptions amount to
demanding that the integral curves of the Killing vector fields provide a
fibration $\pi\colon{\cal M}\to M$ of the spacetime manifold by spacelike
surfaces. The two-dimensional manifold $M$ is the ``space of orbits'' of the
Killing vector fields.

The spacetime metric $g_{\alpha\beta}$ defines three functions
$$
\lambda_{AB}=g_{\alpha\beta}K^\alpha_AK^\beta_B,\tag 22
$$
 representing the lengths and
inner products of the Killing vector fields.  Because the Killing
vectors are spacelike and linearly independent at each point, this
symmetric
matrix will have a symmetric
 inverse $\lambda^{AB}$. We can use $\lambda^{AB}$
to define a projection operator $\gamma^\alpha_\beta$ into the
(co-)tangent space to each point of $M$:
$$
 \gamma^\alpha_\beta=\delta^\alpha_\beta -
\lambda^{AB}K^\alpha_AK_{B\beta}.\tag 23
$$
 The associated tensor field
$$
\gamma_{\alpha\beta}=g_{\alpha\beta}-\lambda^{AB}K_{A\alpha}K_{B\beta},\tag 24
$$
 satisfies
$$
{\cal L}_{{\ss K_A}}\gamma_{\alpha\beta}=0\quad{\rm and}\quad
K^\alpha_A \gamma_{\alpha\beta}=0.\tag 25
$$
There is a one-to-one correspondence
between tensor fields on $M$ and tensor fields on $\cal M$ with
vanishing Lie derivatives along $K^\alpha_A$ and which are completely
``orthogonal'' to $K^\alpha_A$ \refto{Geroch1971a,Geroch1971b}.  This
correspondence means that
$\gamma_{\alpha\beta}$
uniquely defines a Lorentz metric on $M$.

Coordinates $x^\alpha=(x^A,x^a)$, $A=1,2$, $a=3,4$ can be chosen such
that
$$
 K_A^\alpha=\delta^\alpha_A,\tag 26
$$
 and the
spacetime metric takes the form
$$
g_{\alpha \beta}dx^\alpha dx^\beta
=\lambda_{AB}dx^Adx^B +\gamma_{ab}dx^adx^b,\tag 27 $$
where
$\lambda_{AB}=\lambda_{AB}(x^a)$, and
$\gamma_{ab}=\gamma_{ab}(x^a)$.

There are only six independent Einstein equations because the orthogonal
transitivity requirement forces
$$
 \gamma^\alpha_\gamma K^\beta_AG_{\alpha\beta}\equiv0.\tag 28
$$
Thus the non-trivial Einstein equations can be taken to be
$$
\eqalign{ K^\alpha_AK^\beta_BG_{\alpha\beta}&=0,\cr
\gamma^\alpha_\gamma\gamma^\beta_\delta G_{\alpha\beta}&=0.}\tag 29
$$
Setting $\lambda=\det(\lambda_{AB})$ these equations can be put into
the following form in the coordinates $(x^A,x^a)$:
$$
\eqalign{
 -{1\over2}\lambda^{1/2}&D^a(\lambda^{-1/2}D_a\lambda_{\ss AB})\cr
&+\lambda_{\ss AB}\big(-{1\over2}{\cal R} + \lambda^{-1/2}D^a D_a \lambda^{1/2}
+{3\over2} D^a\lambda^{-1/2}D_a\lambda^{1/2}
-{3\over8}D^a\lambda^{\ss CD}D_a\lambda_{\ss CD}\big)=0,}\tag 2101
$$
$$
\eqalign{
{1\over4}D_c\lambda^{\ss AB}D_d\lambda_{\ss AB}
-&D_c(\lambda^{-1/2}D_d\lambda^{1/2})\cr
+&\gamma_{cd}\big(\lambda^{-1/2}D^a D_a\lambda^{1/2}
+{1\over2}D^a\lambda^{-1/2}D_a\lambda^{1/2}
-{1\over8}D^a\lambda^{\ss AB} D_a\lambda_{\ss AB}\big)
=0.}\tag 2102
$$
Here
$D_a$ is the derivative operator on $M$ compatible with $\gamma_{ab}$
and ${\cal R}$ is the scalar curvature of $D_a$.  Indices for
tensor fields on $M$ (Latin lowercase) are
lowered and raised with $\gamma_{ab}$
and its inverse $\gamma^{ab}$.  It will be useful later
to note that eq. \(2102) implies
$$
 D^aD_a(\lambda^{1\over2})=0.\tag 211
$$
All of our considerations will be formulated on the space of orbits $M$.  The
Einstein equations \(2101), \(2102) are viewed as a generally covariant
system of partial differential equations for ``matter fields''
$\lambda_{AB}$ and Lorentz metric $\gamma_{ab}$ on the two-dimensional
manifold $M$.

\bigskip \line{\bf 2a. Cylindrical symmetry.\hfill}

Our model for open universes is defined by taking ${\cal M}={\bf R}^3\times
 {\bf S}^1$
and ${\cal G}={\bf S}^1\times {\bf R}$. Note that $\cal M$ is diffeomorphic to
the manifold obtained by removing a timelike 2-plane
(swept out by, say, the z-axis) {}from
${\bf R}^4$.  The Killing coordinates are
denoted by $x^A=(\phi,z)$, where $\phi\in(0,2\pi)$ and $z\in
(-\infty,\infty)$.  The space of orbits is $M={\bf R}\times {\bf R}^+$,
where ${\bf R}^+$ is the manifold of positive-definite real numbers.
We define $R:= \lambda^{1\over2}$ and demand that $\nabla_\alpha R$ is
everywhere spacelike.  We will call spacetimes with these properties
{\it cylindrically symmetric}.  In what follows it will be useful to
employ coordinates on $M$ that are adapted to a foliation by spacelike curves
 ${\bf R}^+\hookrightarrow M$.  These coordinates will be denoted
by $x^a=(t,r)$
where
$t\in(-\infty,\infty)$ and $r\in(0,\infty)$.  The radial variable $r$
will be restricted by the requirement that ${\partial R\over\partial
r}>0$.  $R$ itself can serve as a radial coordinate on ${\bf R}^+$.
Because of \(211), $R$ will be a harmonic coordinate when the Einstein
equations hold.  The harmonic conjugate $T$ to $R$ is defined by
$$
D_aT=\epsilon_{a}^{\phantom{a}b}D_bR,\tag 212
$$
 where $\epsilon_{ab}$
is the volume form on $M$ defined by $\gamma_{ab}$.  The coordinates
$t=T$ and $r=R$ form a conformal coordinate chart when the Einstein
equations hold, that is, in these coordinates the metric on $M$ is given
by
$$
 ds^2=\Omega^2(-dT^2 + dR^2),\tag 213
$$
 where $\Omega=\Omega(T,R)$
is a positive-definite function.

We parametrize $\lambda_{AB}$ in terms of scalar fields $(R,
\psi,\psitilde)$ via
$$
 \lambda_{AB}dx^Adx^B=R^2e^{-\psi}d\phi^2 +
e^\psi(dz+\psitilde d\phi)^2.\tag 214
$$
In preparation for the
Hamiltonian formulation of the Einstein equations \(2101) and \(2102), we
foliate $M$ with
spacelike curves $t= const$ and parametrize the metric on $M$
via
$$
 \gamma_{ab}dx^adx^b=[-(N^{\perp})^2 + e^{\gamma-\psi}(N^r)^2]dt^2
+2e^{\gamma-\psi}N^rdtdr + e^{\gamma-\psi}dr^2.\tag 215
$$
 As tensor
fields on the curves $t=const$, $N^\perp$ is a scalar field called the ``lapse
function'', $N^r$ is a vector field called the ``shift vector'' (equivalent
to a
density of weight minus one on ${\bf R}^+$), and
$\gamma$ is the natural logarithm of a covariant rank-2 tensor field
(equivalent to the logarithm of a density of weight-two).
The functions $(R, \psi,\psitilde)$ are scalar fields on $t=
const$.  All the fields $(N^\perp, N^r, \gamma, R, \psi, \psitilde)$
are functions of the coordinates $(t,r)$ only; see the Appendix for the
boundary and fall-off conditions we use for these fields.  In terms of
this parametrization, the conformal factor $\Omega$ arising in the coordinates
$x^a=(T,R)$ (see \(213)) is given by
$$
\Omega=e^{(\gamma-\psi)/2}.\tag 216
$$

It is worth noting some special cases of the parametrization given
above. If the Killing vectors are each hypersurface orthogonal, it follows
that $\psitilde=0$, {\it i.e.}, the matrix $\lambda_{AB}$ is diagonal,
indicating orthogonality of the Killing vectors.  This special case of
cylindrical symmetry in which each $K^\alpha_A$ is hypersurface orthogonal
will be called {\it whole cylindrical symmetry}. The solutions to the field
equations possessing whole cylindrical
symmetry are the well-known Einstein-Rosen waves.  Kucha\v r
assumes
whole cylindrical symmetry in his analysis \refto{Kuchar1971} and, with
$\psitilde=0$,
our parametrization of the spacetime metric is identical to his.
When
$$
 N^\perp=1,\quad N^r=0,\quad\gamma=0,\quad
R=r,\quad\psi=0,\quad\psitilde=0\tag 217
$$
the spacetime is flat and
the metric is expressed in cylindrical coordinates.

Let an overdot denote differentiation with respect to $t$.  The field
equations \(2101), \(2102) in the parametrization \(214), \(215) can be
obtained from the following Hamiltonian form of the action:
$$
 \eqalign{
S[N^\perp,N^r,\gamma,R,\psi,\psitilde,\pi_\gamma,
\pi_R,\pi_\psi,\pi_\psitilde] =\int_{t_1}^{t_2}dt\,\int_0^\infty
dr\,&\left(\pi_\gamma\dot\gamma +\pi_R\dot R + \pi_\psi\dot\psi +
\pi_{\psitilde}\dot\psitilde\right)\cr &-\int_{t_1}^{t_2}dt\,{\bf
H},}\tag 218
$$
 where the Hamiltonian is
$$
 {\bf
H}=2N^\perp(\infty)(1-e^{-\gamma(\infty)/2})+ \int_{0}^\infty
dr\,\left(N^\perp{\cal H}_\perp + N^r{\cal H}_r\right),\tag 219
$$
and we have denoted the limits of $N^{\perp}(r)$ and $\gamma(r)$ as
$r\rightarrow\infty$ by $N^{\perp}(\infty)$ and $\gamma(\infty)$.
 The super-Hamiltonian and super-momentum are given by
$$
 \eqalignno{ {\cal
H}_\perp:&=e^{(\psi-\gamma)/2}\Bigg[-
\pi_\gamma\pi_R+2R^{\prime\prime}-
R^\prime\gamma^\prime +{1\over2}\left(R\psi^{\prime2}+R^{-
1}\pi_{\psi}^2\right)\cr
&\qquad\qquad\qquad+{1\over2}\left(R e^{-2\psi}\pi_{\psitilde}^2
+R^{-1}e^{2\psi}\psitilde^{\prime2}\right)\Bigg],&(220)\cr
{\cal H}_r:&=-2\pi_\gamma^\prime
+\pi_\gamma\gamma^\prime+\pi_RR^\prime+\pi_\psi\psi^\prime
+\pi_{\psitilde}\psitilde^\prime,&(221)}
$$
 where a prime indicates a
derivative with respect to $r$.  For each $t$ the momentum variables
$(\pi_\gamma, \pi_R,\pi_\psi, \pi_\psitilde)$ are scalar densities of
weight-one on ${\bf R}^+$.  The Hamiltonian action functional \(218) can
be obtained by (i) expressing the ADM action in terms of the parametrization
\(214) and \(215), (ii) integrating out the Killing coordinates $z$ and $\phi$
(the range of $z$ should be made finite), (iii) dividing the result by
the range
of $z$ and the range of $\phi$.

Extremising the action \(218) with respect to variations of
$\pi_\gamma$, $\pi_R$, $\pi_\psi$, and $\pi_\psitilde$ produces the
definitions of the momenta in terms of the ``velocities'' $\dot\gamma$,
$\dot R$, $\dot\psi$, and $\dot\psitilde$. Extremising the action \(218)
with respect to variations of $N^\perp$, $N^r$, $\gamma$, $R$, $\psi$,
and $\psitilde$, where the asymptotic values of $N^\perp$ and $N^r$ are
held fixed, yields six equations which, given the definitions of the
momenta, are equivalent to the six field equations \(2101), \(2102) in the
parametrization \(214), \(215).  In particular, the field equations
arising {}from varying the lapse and shift are the constraints
$$
 {\cal
H}_\perp\approx0\qquad {\rm and}\qquad {\cal H}_r\approx0,\tag 222
$$
which correspond to the normal-normal and normal-tangential projections
of the field equations \(2102) relative to the curve $t=const$. The
remaining four equations are evolution equations for the spatial metric
variable $\gamma$ and the ``matter fields'' $R$, $\psi$, and
$\psitilde$.

Let us make two remarks on the Hamiltonian variational principle we have
described.  (i) It is important to note that the canonical variables are
restricted by the assumption that $\nabla_\alpha R$ is spacelike on $\cal M$,
which implies that $D_aR$ is spacelike on $M$.  In
Hamiltonian form, this restriction is
$$
 R^\prime>|\pi_\gamma|.\tag 223
$$
(ii) The boundary term at infinity that appears in the Hamiltonian is
needed to render the action and Hamiltonian functionally differentiable
with the boundary conditions given in the Appendix.  On solutions to the
field equations the Hamiltonian is given by the boundary term, which we
identify as the energy generating time evolution at infinity
characterized by $N^\perp(\infty)$.   Note that our boundary conditions
are such that $N^\perp(\infty)$ is fixed,
that is, not subject to variation.  This allows us to add constant
multiples of $N^\perp(\infty)$ to the surface term without spoiling the
differentiability of the action or Hamiltonian.  We have used this
freedom to normalize the Hamiltonian so that it vanishes when spacetime
is flat, which occurs when $\gamma(\infty)=0$.  The energy associated
with time translations described by unit lapse function at infinity is
$$
 E=2(1-e^{-\gamma(\infty)/2}).\tag 224
$$
It follows {}from the field
equations that $\gamma(\infty)$, known
as the ``C-energy'', is conserved and non-negative (see, {\it e.g.},
\refto{Berger1995}). Hence $E$ is bounded
{}from below and is minimized on flat spacetime. Note also that $E$ is bounded
from above, which is due to the geometric interpretation of $E$ as a deficit
angle (divided by $\pi$) at infinity.  See \refto{Ashtekar1994} for a
detailed discussion of this interesting behavior of the energy of
cylindrically symmetric spacetimes.

To summarize, cylindrically symmetric spacetimes constitute a
constrained Hamiltonian system.  The phase space $\Gamma$ is the space
of fields $(\gamma, R, \psi, \psitilde, \pi_\gamma, \pi_R,\pi_\psi,
\pi_\psitilde)$ on ${\bf R}^+$ with boundary conditions as specified in
the Appendix, and with the restriction \(223). The action functional
\(218) defines the symplectic 2-form $\Omega$ on $\Gamma$.
$\Omega$ maps a pair of tangent vectors
$$
 X=(\delta\gamma, \delta R,
\delta\psi, \delta\psitilde, \delta\pi_\gamma,
\delta\pi_R,\delta\pi_\psi, \delta\pi_\psitilde)\tag 225
$$
 and
$$
\widehat X=(\widehat\delta\gamma, \widehat\delta R, \widehat\delta\psi,
\widehat\delta\psitilde, \widehat\delta\pi_\gamma,
\widehat\delta\pi_R,\widehat\delta\pi_\psi, \widehat\delta\pi_\psitilde)
\tag 226
$$
 to the real number
$$
 \Omega(X,\widehat X)=\int_0^\infty
dr\, \left(\delta\pi_\gamma\widehat\delta\gamma
+\delta\pi_R\widehat\delta R +\delta\pi_\psi\widehat\delta\psi
+\delta\pi_\psitilde\widehat\delta\psitilde
-[\delta\longleftrightarrow\widehat\delta]\right).\tag 227
$$
{}From
\(227) it follows that $(\gamma, R, \psi, \psitilde)$ and $(\pi_\gamma,
\pi_R,\pi_\psi, \pi_\psitilde)$ are, respectively, canonical coordinates
and momenta for $\Gamma$ and hence satisfy the canonical Poisson
bracket
relations, {\it e.g.},
$$
\{\gamma(r),\pi_\gamma(\bar r)\}=\delta(r,\bar r).\tag22701
$$
In terms of the Poisson bracket $\{\cdot,\cdot\}$, the time
evolution of a function $F\colon\Gamma\to {\bf R}$ is given by
$$
 \dot
F=\{F,{\bf H}\}.\tag 228
$$
For any choice of lapse and shift,
dynamical evolution takes place entirely on
the constraint surface $\overline\Gamma\hookrightarrow\Gamma$ defined
by
the constraints \(222).  This follows {}from the fact that the constraints
are ``first class''. More precisely, the Poisson algebra of the
super-Hamiltonian and super-momentum is the algebra of deformations of
spacelike curves in $M$ \refto{Teitelboim1984}.

In two-dimensional spacetimes, such as we have here, it is
convenient to work with a rescaled super-Hamiltonian which is a scalar
density of weight-two.  We define the weight-two super-Hamiltonian via
$$
 {\cal H}=e^{(\gamma-\psi)/2}{\cal H}_\perp.\tag 229
$$
 Of course the
super-Hamiltonian constraint ${\cal H}_\perp\approx0$ is equivalent to
${\cal H}\approx0$, and we can write
$$
 N^\perp{\cal H}_\perp=N{\cal
H},\tag 230
$$
where $N$ is a scalar density of weight $-1$, which is equivalent to a vector
in
one dimension, obtained by
$$
 N=e^{(\psi-\gamma)/2}N^\perp.\tag 231
$$
We can now
vary the action
$$
 \eqalign{ \bar
S[&N,N^r,\gamma,R,\psi,\psitilde,\pi_\gamma,\pi_R,\pi_\psi,\pi_\psitilde
]\cr&=\int_{t_1}^{t_2}dt\,\int_0^\infty
dr\,\left(\pi_\gamma\dot\gamma+\pi_R\dot R + \pi_\psi\dot\psi +
\pi_{\psitilde}\dot\psitilde -N{\cal H} - N^r {\cal H}_r\right)
-\int_{t_1}^{t_2}dt\,N(\infty)\gamma(\infty),}\tag 232
$$
 with respect
to its arguments and obtain equations still equivalent to \(2101), \(2102) once
the correspondence \(231) is made.  In this variational principle we
hold fixed the asymptotic value of $N$.   When using the
weight-two super-Hamiltonian and associated lapse density the
Hamiltonian is given by
$$
 \bar{\bf H}=N(\infty)\gamma(\infty)+
\int_{0}^\infty dr\,\left(N{\cal H} +N^r{\cal H}_r\right),\tag 233
$$
where the surface term is again chosen to make $\bar S$ and $\bar {\bf
H}$ differentiable with our boundary conditions and to yield $\bar{\bf H}=0$
when $\gamma(\infty)=0$. Evidently, the C-energy is associated with time
translations defined by unit lapse {\it density} at infinity.

The advantage of the weight-two super-Hamiltonian is that the Poisson
algebra of constraints is now a Lie algebra \refto{liealgebranote}.   In
detail, define
smeared constraints:
$$
 {\cal H}(N)=\int_0^\infty dr\, N {\cal H}
\qquad{\rm and}\qquad {\cal H}_r(N^r)=\int_0^\infty dr\, N^r {\cal
H}_r,\tag 234
$$
 where $N^r$ is a given vector field and $N$ is a given
scalar density of weight $-1$.   Direct computation then shows that
$$
\eqalign{
\{{\cal H}(N), {\cal H}(M)\} &= {\cal H}_r(J^r),\cr
\{{\cal H}(N),{\cal H}_r(M^r)\} &= {\cal H}(K),\cr
\{{\cal H}_r(N^r), {\cal H}_r(M^r)\} &={\cal H}_r(L^r),}\tag 235
$$
where
$$
 J^r=NM^\prime -MN^\prime, \quad
K=NM^{r\prime}-M^rN^\prime, \quad L^r=N^rM^{r\prime}-
M^rN^{r\prime}.\tag
236
$$

\bigskip \line{\bf 2b. Toroidal symmetry.\hfill}

Our model for closed universes is obtained by setting
${\cal M}={\bf R}^+\times {\bf T}^3$ and choosing the isometry group to be
${\cal G}={\bf T}^2$ with coordinates $y,z\in(0,2\pi)$.
The space of orbits is a cylinder, $M={\bf R}^+\times {\bf S}^1$.
We are considering one of the Gowdy models
\refto{Gowdy1974},
and we shall call spacetimes with the above properties {\it toroidally
symmetric}.   In these spacetimes we define $\tau:=\lambda^{1\over2}$.
It can be shown  that toroidally symmetric vacuum spacetimes are foliated
by spacelike surfaces whose leaves are defined by $\tau=const$
\refto{Gowdy1974,Moncrief1980}.
Hence the gradient of $\tau$ is timelike, which we will
assume in all that follows.  According to observers ``at rest'' relative
to the slices $\tau=const$, the toroidally symmetric spacetimes
expand forever {}from a ``big bang'' at $\tau=0$.  Coordinates on $M$ that
are adapted to a foliation by spacelike circles will be denoted $(t,x)$,
where $t\in(0,\infty)$ and $x\in(-\infty,\infty)$ with the
identification $x\sim x + 2\pi$.  We will demand that a time coordinate $t$
on $M$ satisfies ${\partial \tau\over\partial t}>0$.   For any solution
of the Einstein equations, $\tau$ is a harmonic time coordinate.  The
harmonic conjugate to $\tau$ is denoted by $X$, which satisfies
$$
 D_a X={\epsilon_a^{\phantom{a}b}D_b\tau\over
{1\over2\pi}\int_{{\ss {\bf S}^1}} ds^a\,\epsilon_a^{\phantom{a}b}D_b\tau},
\tag 237
$$
 where $ds^a$ is the oriented line element on a spacelike circle.
Because $D_a\tau$ is timelike, the denominator in \(237) never vanishes.  By
virtue of \(211), the integral in the denominator is independent of the choice
of
spacelike circle, {\ie} is a constant of motion.   This integral is introduced
so that on any circle $t=const$
we have that
$$
 X(x+2\pi)-X(x) = 2\pi.\tag 238
$$
 With the
identification $X\sim X+2\pi$, the coordinates $t=\tau,\ x=X$ are
adapted to a spacelike foliation of $M$; in these coordinates the metric
on $M$ is of the form
$$
 ds^2=\Omega^2(-d\tau^2 + dX^2).\tag 239
$$

The Hamiltonian formulation of toroidally symmetric spacetimes closely
parallels that obtained for cylindrically symmetric spacetimes.
Essentially, the toroidally symmetric case differs by the change of
notation $R\leftrightarrow \tau$ and the fact that the Cauchy surfaces
in $M$ are now compact circles parametrized by $x$ instead of
non-compact half-lines parametrized by $r$.  We parametrize
$\lambda_{AB}$ in terms of scalar fields $(\tau, \psi,\psitilde)$ via
$$
\lambda_{AB}dx^Adx^B=\tau^2e^{-\psi}dy^2 + e^\psi(dz+\psitilde
dy)^2.\tag 240
$$
The metric on $M$ is parametrized relative to an
arbitrary foliation by spacelike circles exactly as in \(215):
$$
\gamma_{ab}dx^adx^b=[-(N^{\perp})^2 + e^{\gamma-\psi}(N^x)^2]dt^2
+2e^{\gamma-\psi}N^xdtdx + e^{\gamma-\psi}dx^2.\tag 241
$$
In the coordinates $x^a=(\tau,X)$ the conformal factor in \(239) is
$$
\Omega=e^{(\gamma-\psi)/2}.\tag24101
$$

When $\psitilde=0$ our parametrization corresponds to a ``polarized
Gowdy model'' \refto{Gowdy1971}, which can be considered the closed universe
analog of the
Einstein-Rosen waves.  When
$$
 N^\perp=1,\quad N^x=0,\quad\gamma=0,\quad
\tau=t,\quad\psi=0,\quad\psitilde=0\tag 242
$$
 the spacetime is flat.
This spacetime can be obtained from 4-dimensional Minkowski spacetime
${\bf M}^4$ as follows. Let
$(\hat t, \hat x, \hat y, \hat z)$ be inertial coordinates on ${\bf M}^4$.
Denote by $I^+$ the chronological future of the origin
in ${\bf M}^4$.  Let $\tilde I$ denote the manifold obtained {}from $I^+$ by
the identifications
$$
 \hat x\sim \hat x+2\pi,\qquad \tanh^{-1}({\hat
y\over\hat t})\sim \tanh^{-1}({\hat y\over\hat t}) + 2\pi,\qquad \hat
z\sim \hat z + 2\pi. \tag 243
$$
 The Minkowski metric $\eta_{\alpha\beta}$
projects to a flat metric on $\tilde I$.  The mapping $\phi\colon {\cal
M}\to \tilde I$ defined by
$$
 \eqalign{ \hat t&=t\cosh y,\cr \hat x&=x,\cr
\hat y&=t\sinh y,\cr \hat z&=z,}\tag 244
$$
 is a diffeomorphism which
identifies the metric on $\cal M$ defined by \(242),
$$
ds^2=-dt^2 + dx^2 + t^2 dy^2 + dz^2,\tag24401
$$
with the flat metric
induced on $\tilde I$.

In terms of a spacelike foliation of $M$ with adapted coordinates $(t,x)$, the
Hamiltonian form of the action is
$$
\eqalign{
S[N^\perp,N^r,\gamma,\tau,\psi,\psitilde,\pi_\gamma,\pi_\tau,\pi_\psi,
\pi_\psitilde] =&\int_{t_1}^{t_2}dt\,\int_0^{2\pi}
dx\,\left(\pi_\gamma\dot\gamma +\pi_\tau\dot \tau + \pi_\psi\dot\psi +
\pi_{\psitilde}\dot\psitilde\right)\cr
 -&\int_{t_1}^{t_2}dt\,{\bf H},}
\tag 245
$$
 where the Hamiltonian is
$$
 {\bf H}=\int_{0}^{2\pi}
dx\,\left(N^\perp{\cal H}_\perp + N^x{\cal H}_x\right).\tag 246
$$
 The
super-Hamiltonian and super-momentum are given by
$$
 \eqalignno{ {\cal
H}_\perp:&=e^{(\psi-\gamma)/2}\left[-
\pi_\gamma\pi_\tau+2\tau^{\prime\prime}-
\tau^\prime\gamma^\prime
+{1\over2}\left(\tau\psi^{\prime2}+\tau^{-1}\pi_{\psi}^2 \right)
+{1\over2}\left(\tau e^{-2\psi}\pi_{\psitilde}^2
+\tau^{-1}e^{2\psi}\psitilde^{\prime2}\right)\right]\cr &\approx0,&(247)\cr
{\cal H}_x:&=-2\pi_\gamma^\prime
+\pi_\gamma\gamma^\prime+\pi_\tau\tau^\prime+\pi_\psi\psi^\prime
+\pi_{\psitilde}\psitilde^\prime\cr &\approx0,&(248)}
$$
 where a prime
denotes differentiation with respect to $x$.  As in the cylindrically
symmetric case, the action functional \(245) can be obtained by
expressing the usual ADM action in the field parametrization \(240) and \(241),
integrating out
the coordinates $y$ and $z$, and dividing by the ranges of these coordinates.
Extremising \(245) with respect to variations of its arguments
leads to the Einstein field equations \(2101) and \(2102)
in our chosen parametrization.

To summarize, points in the phase space $\Gamma$ of toroidally
symmetric
spacetimes are defined by the smooth tensor fields on the circle
$(\gamma, \tau, \psi, \psitilde, \pi_\gamma, \pi_\tau,\pi_\psi,
\pi_\psitilde)$.  These fields are restricted by the requirement that
$\nabla_\alpha\tau$ is timelike on $\cal M$, which means
$D_a\tau$ is timelike on $M$; in Hamiltonian form this requirement is
$$
\pi_\gamma<-|\tau^\prime|.\tag 249
$$
 The action \(245) defines the
symplectic structure on $\Gamma$.  The symplectic 2-form acting on a
pair of tangent vectors to $\Gamma$ is given by
$$
 \Omega(X,\widehat
X)=\int_0^{2\pi} dx\, \left(\delta\pi_\gamma\widehat\delta\gamma
+\delta\pi_\tau\widehat\delta \tau +\delta\pi_\psi\widehat\delta\psi
+\delta\pi_\psitilde\widehat\delta\psitilde
-[\delta\longleftrightarrow\widehat\delta]\right),\tag 250
$$
 so that,
with $R\leftrightarrow \tau$, the canonical coordinates and momenta are
as before.  The weight-two super-Hamiltonian is defined via \(229), and
the Poisson-algebraic properties of the constraint functions are as in
the cylindrically symmetric case.  Dynamical evolution takes place on
the constraint surface $\overline\Gamma$ defined by the constraints \(247)
and \(248), and is generated by the Hamiltonian \(246).

\taghead{3.}\bigskip \line{\bf 3.  Canonical Transformations.\hfill}

In this section we will exhibit canonical transformations {}from slight
extensions of the gravitational phase spaces of \S2 to phase spaces for
parametrized field theories on a fixed background spacetime.  The
extensions are needed because the gravitational phase space
is not quite adequate to define embeddings of hypersurfaces into
Ricci-flat spacetimes.  This difficulty arises for the full theory in both
closed universes \refto{CGT1992a}
and open universes \refto{Regge1974}, and can be considered a
``global
problem of time'' \refto{Kuchar1992}.   While it is possible to phrase all of
our
results directly on the four-dimensional spacetime manifold $\cal M$, it
is far more convenient to express our results on the space of orbits
$M$, and we will present our analysis on this effective
two-dimensional spacetime manifold.

Before specializing to our two Killing vector models, let us outline the
basic strategy for the full theory \refto{Kuchar1972, Kuchar1992}.  Denote the
usual gravitational
phase space variables by $(q_{ij},p^{ij})$ and the super-Hamiltonian and
super-momentum by $({\cal H}_\perp, {\cal H}_i)$.  We seek a canonical
transformation
$$
 (q_{ij},p^{ij})\longrightarrow
(X^\alpha,\Pi_\alpha,q^{\ss\bf A},p_{\ss\bf A}),\tag 2501
$$
where $\alpha=0,1,2,3$ and
${\bf A}=1,2$ such that, on solutions to the equations of motion and
constraints, $X^\alpha\colon\Sigma\to {\cal M}$ represents a spacelike
embedding of a Cauchy surface $\Sigma$ into the spacetime manifold $\cal
M$.  The transformation must allow the constraints ${\cal
H}_\perp\approx0\approx {\cal H}_i$ to be resolved for the momenta
conjugate to the embeddings, {\it i.e.}, in the new variables the
constraints are equivalent to
$$
 H_\alpha:=\Pi_\alpha +
h_\alpha(X^\alpha,q^{\ss\bf A},p_{\ss\bf A})\approx0.\tag 251
$$
This formulation of the
canonical theory has the following interpretation.
The gravitational variables encoded in $X^\alpha$ are used to identify
instants of time and points of space at which the true degrees of
freedom $(q^{\ss\bf A},p_{\ss\bf A})$ are being measured.  The embeddings are
``pure
gauge'', {\it i.e.}, arbitrary; their conjugate momenta are
completely determined in terms of the embeddings and true degrees of
freedom by the constraints \(251).  The constraint functions $H_\alpha$,
when integrated against functions $N^\alpha$, generate the dynamical
evolution of the true degrees of freedom $(q^{\ss\bf A},p_{\ss\bf A})$ as the
embedding they are on is
deformed, via $\delta X^\alpha=N^\alpha$, through the Ricci-flat
spacetime for which $(q^{\ss\bf A},p_{\ss\bf A})$ are Cauchy data. The
densities of weight-one $h_\alpha(X^\alpha,q^{\ss\bf A},p_{\ss\bf A})$
represent
the energy-momentum current
of
$(q^{\ss\bf A},p_{\ss\bf A})$ through the hypersurface embedded by $X^\alpha$.
Gravitational dynamics on phase space is thus cast into the form of a
``parametrized field theory'' on $\cal M$.

\bigskip \line{\bf 3a. Open universes: cylindrical symmetry.\hfill}

As we saw in \S2a, the metric variable $R$ and its harmonic conjugate
$T$ define a conformal coordinate chart on cylindrically symmetric
spacetimes (reduced to $M$).  We can thus define a spacelike curve ${\bf
R}^+\hookrightarrow M$ by giving its parametric description
$(T(r),R(r))$.  In the canonical formalism we can therefore view the
phase space variable $R(r)$ as one part of an embedding of ${\bf R}^+$
into $M$.  To complete the definition of the embedding we must express
$T$ as a function on the phase space, which will be again denoted $T(r)$.
Using the pullback of \(212) to a spacelike curve $t=const$ and
the Hamilton equations
we find
$$
 T^\prime = -\pi_\gamma.\tag 252
$$
 This equation can be
integrated to give
$$
 T(r) = T(\infty)-\int_\infty^rd\bar r\,
\pi_\gamma(\bar r).\tag 253
$$
Because of the restriction \(223), the
variables $(T(r), R(r))$ define a spacelike embedding on solutions of
the equations of motion.  Unfortunately, given a point in the
gravitational phase space $\Gamma$, the embedding is not uniquely
specified because the gravitational data do not fix the value of the
integration constant $T(\infty)$, which represents the asymptotic
location of the embedded curve.  This difficulty is not a consequence of
our use of the conformal coordinates $(T,R)$ to specify an embedding,
but is a general feature of general relativity of open universes.  For
example, in \refto{Regge1974} the Hamiltonian formulation of general relativity
in
asymptotically flat universes is considered, and it is shown that the
geometrodynamical data must be supplemented by a finite number of
additional degrees of freedom in order to determine a spacelike
hypersurface.  To implement this idea in the context of cylindrically
symmetric spacetimes, we will extend the phase space $\Gamma$ of \S2a.  We then
make a canonical change of variables on the extended phase space in which
certain degrees of freedom are identified as spacelike embeddings.  The
resulting Hamiltonian structure of the model is then interpreted
as a generally
covariant form of a theory of harmonic maps by a sequence of canonical
transformations on the appropriate harmonic map phase space. We begin by
finding
the appropriate extended phase space for cylindrically symmetric spacetimes.

{}From the definition of the lapse density $N$ and the time $T$, or
equivalently, from the equations of motion for $T$, it follows
that
the rate of change of $T(\infty)$ with respect
to $t$ is
the asymptotic value of the lapse density $N(\infty)$.
  This motivates the following construction.  We introduce a new
degree of freedom, $\tau_\infty$, which is the time displayed by a clock
at infinity that measures time $T$.   The lapse density at infinity is
expressed as
$$
 N(\infty)=\dot\tau_\infty.\tag 254
$$
If we insist on
keeping fixed the asymptotic value of the lapse density, then we also
keep fixed $\dot\tau_\infty$, and nothing is changed except notation.
However, we can treat $\tau_\infty$ as a new dynamical variable which is
to be varied in the action.  This is the usual logic of the
parametrization process as applied to the ``point at infinity''. If we
parametrize at infinity, the Hamiltonian form of the action \(232) can
be written as
$$
 \eqalign{ &\tilde
S[N,N^r,\gamma,R,\psi,\psitilde,\pi_\gamma,\pi_R,\pi_\psi,\pi_\psit,
\tau_\infty]\cr
&=\int_{t_1}^{t_2}dt\,\int_0^\infty
dr\,\left(\pi_\gamma\dot\gamma +\pi_R\dot R + \pi_\psi\dot\psi +
\pi_{\psitilde}\dot\psitilde -N{\cal H} - N^r{\cal H}_r\right)-
\int_{t_1}^{t_2}dt\, \dot\tau_\infty\gamma(\infty).}\tag 255
$$
 By
adding the new variable $\tau_\infty$ to the Hamiltonian action
principle we get additional equations that, however, are
equivalent to the original equations.  In detail, by varying
$\tau_\infty$ we obtain conservation of the C-energy,
$\dot\gamma(\infty)=0$, which already followed {}from the other equations
of motion and so does not alter the content of the field equations.  The
other new equation comes {}from varying $\gamma$.  Prior to
parametrizing
at infinity, the variation of $\gamma$ led to one of the field
equations \(2101), \(2102), and a potential boundary equation was eliminated
by the boundary term in the Hamiltonian.  After parametrization, the
boundary equation survives and yields equation \(254), which recovers the
desired definition of $\tau_\infty$.

To summarize, we can enlarge the phase space $\Gamma$ of cylindrically
symmetric spacetimes by adding a single variable $\tau_\infty$.  The
extended phase space will be denoted $\Gamma^\star$.  The extrema of
the action functional \(255) still define cylindrically symmetric vacuum
spacetimes, but now in terms of the extended set of variables.  By
enlarging the phase space in this manner we are able to define the
asymptotic location of spatial curves using dynamical variables.  The
action $\tilde S$, however, is not in Hamiltonian form because the surface
term now enters as a ``kinetic term'' and destroys the canonical nature
of the phase space coordinates $\gamma$ and $\pi_\gamma$.  We still must
find canonical coordinates and momenta on $\Gamma^\star$.  Indeed,
we
must show that $\Gamma^\star$ is a symplectic manifold.  We will take
care of these issues, while at the same time providing the cylindrically
symmetric version of the canonical transformation \(2501), in the
following.

Let us define the phase space $\Upsilon$ for a field theory on
$M$ as follows.  A point in phase space is defined by the functions
$(T,R,\psi,\psitilde,\Pi_T,\Pi_R,\pi_\psi,\pi_\psitilde)$ on ${\bf
R}^+$, where $(T,R,\psi,\psitilde)$ are scalar functions and
$(\Pi_T,\Pi_R,\pi_\psi,\pi_\psitilde)$ are scalar densities of weight-one.
Boundary and fall-off conditions on these scalar fields and scalar
densities are as indicated in the Appendix.  We will need some restrictions
on the functions $T(r)$ and $R(r)$ so that they can be interpreted as spacelike
embeddings of ${\bf R}^+$ into $M$. For reasons which will be clearer in a
moment,
we demand
$$
R^\prime>|T^\prime|,\tag 256
$$
and we include this inequality in the
definition of $\Upsilon$.  We define a symplectic 2-form, $\Xi(X,\widehat X)$
on $\Upsilon$ by its
action on a pair of vectors
$$
 X=(\delta T,\delta R,\delta \psi,\delta
\psitilde,\delta \Pi_T,\delta \Pi_R,\delta \pi_\psi,\delta
\pi_\psitilde)\tag 257
$$
 and
$$
 \widehat X=(\widehat \delta T,\widehat
\delta R,\widehat \delta \psi,\widehat \delta \psitilde,\widehat \delta
\Pi_T,\widehat \delta \Pi_R,\widehat \delta \pi_\psi,\widehat \delta
\pi_\psitilde)\tag 258
$$
 at a point of $\Upsilon$.  The symplectic form
is defined by
$$
 \Xi(X,\widehat X)=\int_0^\infty dr\, \left(\delta
\Pi_T\widehat\delta T +\delta \Pi_R\widehat\delta R + \delta \pi_\psi
\widehat\delta \psi +\delta\pi_\psitilde\widehat\delta \psitilde
- [\delta\longleftrightarrow\widehat\delta]\right).\tag 259
$$
 For the moment, the symplectic manifold
$(\Upsilon,\Xi)$ is to be viewed as logically independent of the
gravitational phase space, that is, $\Upsilon$ is being thought of as
simply a space of functions
$(T,R,\psi,\psitilde,\Pi_T,\Pi_R,\pi_\psi,\pi_\psitilde)$ upon which we
have defined a symplectic structure.  The symplectic structure is
defined so that
$(T,R,\psi,\psitilde)$ and $(\Pi_T,\Pi_R,\pi_\psi,\pi_\psitilde)$
are, respectively, canonical
coordinates and momenta.

Now consider the following map {}from $\Gamma^\star$ to $\Upsilon$.
As
suggested by our notation, we will identify the variables
$R,\psi,\psitilde,\pi_\psi,\pi_\psitilde$ in $\Gamma^\star$ and
$\Upsilon$. The remaining portion of the map is defined by
$$
\eqalignno{ T(r) &= \tau_\infty-\int_\infty^rd\bar r\,\pi_\gamma(\bar
r),&(260)\cr
\Pi_T&=-\gamma^\prime + \left[\ln(R^{\prime2}-
\pi_\gamma^2)\right]^\prime,&(261)\cr \Pi_R&=\pi_R +
\left[\ln\left({R^\prime-\pi_\gamma\over
R^\prime+\pi_\gamma}\right)\right]^\prime.&(262)}
$$
 Note that this
transformation is consistent with $\Pi_T$ and $\Pi_R$ being scalar
densities of weight-one.  This map can be inverted; the relevant
formulas are
$$
 \eqalignno{ \tau_\infty&=T(\infty),&(263)\cr
\gamma(r)&=\ln(R^{\prime2}-T^{\prime2})-\int_0^rd\bar r\, \Pi_T(\bar
r),&(264)\cr \pi_\gamma&=-T^\prime,&(265)\cr
\pi_R&=\Pi_R-\left[\ln\left({R^\prime+T^\prime\over R^\prime-
T^\prime}\right)\right]^\prime.&(266)}
$$
 The transformation
\(260)--\(262) is a diffeomorphism which identifies $\Gamma^\star$ and
$\Upsilon$.  In particular, the inequality \(256) is precisely the
restriction \(223) on $\Gamma^\star$, and is needed for the
transformation to be well-defined.  We can use this diffeomorphism to
express the action \(232) as a functional $S^\star$ of curves in $\Upsilon$:
$$
 \eqalign{
S^\star[N,N^r,T,R,&\psi,\psitilde,\Pi_T,\Pi_R,\pi_\psi,\pi_\psitilde]\cr
&=\int_{t_1}^{t_2}dt\,\int_0^\infty dr\,\left(\Pi_T\dot T + \Pi_R\dot R
+\pi_\psi\dot\psi + \pi_\psitilde\dot\psitilde -N H -N^rH_r\right),}\tag
267
$$
 where
$$
 \eqalignno{ H&=\Pi_TR^\prime +\Pi_RT^\prime
+{1\over2}\left(R\psi^{\prime2}+R^{-1}\pi_{\psi}^2\right)
+{1\over2}\left(Re^{-2\psi}\pi_{\psitilde}^2
+R^{-1}e^{2\psi}\psitilde^{\prime2}\right)\approx0,&(268)\cr
H_r&=\Pi_TT^\prime
+\Pi_RR^\prime + \pi_\psi\psi^\prime
+\pi_{\psitilde}\psitilde^\prime\approx0.&(269)}
$$
 Note that the
surface term contribution to the action has dropped out.  Indeed,
{}from this action we see that the variables
$(T,R,\psi,\psitilde)$ and $(\Pi_T,\Pi_R,\pi_\psi,\pi_\psitilde)$
are canonical
coordinates and momenta for the phase space $\Gamma^\star$.  We
have (i) shown that $\Gamma^\star$ is a symplectic manifold by
exhibiting a diffeomorphism {}from $\Gamma^\star$ to the symplectic
manifold $(\Upsilon,\Xi)$, and (ii) exhibited a canonical coordinate
chart $(T,R,\psi,\psitilde,\Pi_T,\Pi_R,\pi_\psi,\pi_\psitilde)$ on
$\Gamma^\star$.

The action $S^\star$ has a nice mathematical interpretation in terms of a
parametrized field theory formulation of harmonic maps on a flat
spacetime, and we shall now spend a little time developing this
interpretation. Recall that harmonic maps $\varphi^A\colon {\bf M}\to {\bf L}$
are
fields on a spacetime $({\bf M},g)$ taking values in a Riemannian manifold
$({\bf L},\sigma)$ and which extremise the ``energy integral''
$$
I[\varphi^A]=-{1\over2}\int_{\bf M}\sqrt{-g}g^{\alpha
\beta}\sigma_{AB}(\varphi)\varphi^A_{,\alpha}\varphi^B_{,\beta},\tag
270
$$
 where
$g_{\alpha\beta}$ is the metric on $\bf M$, and $\sigma_{AB}$ is the
metric on $\bf L$.  Note that we are using capital Latin indices to label the
harmonic maps; in this discussion these indices should not be confused with
those labeling the Killing vectors used to define the gravitational model.
Also, the spacetime manifold ${\bf M}$ and metric $g_{\alpha\beta}$ for the
harmonic map theory should not be confused with the gravitational ${\cal M}$
and $g_{\alpha\beta}$.

Introduce a spacelike foliation $X\colon{\bf R}\times\Sigma\to {\bf M}$
characterized by
lapse $N^\perp$ and shift $N^i$.  The Hamiltonian form of the action
\(270) is
$$
 I[\varphi^A,\pi_A]=\int_{{\bf
R}\times\Sigma}\left(\pi_A\dot\varphi^A-N^\perp h_\perp - N^i
h_i\right),\tag 271
$$
 where lowercase Latin indices denote tensors on $\Sigma$, and the energy
and momentum densities are
$$
 \eqalignno{
h_\perp&={1\over2}\left({1\over\sqrt{q}}\sigma^{AB}\pi_A\pi_B +
\sqrt{q}q^{ij}\sigma_{AB}\varphi^A_{,i} \varphi^B_{,j}\right),&(272)\cr
h_i&=\pi_A\varphi^A_{,i}.&(273)}
$$
Here $q_{ij}$ is the induced metric
on each hypersurface $\Sigma$ of the foliation.  Variation of
$I[\varphi^A,\pi_A]$ with respect to $\varphi^A$ and $\pi_A$ yields equations
equivalent to those obtained by varying \(270). At this point the
spacetime metric and foliation, while arbitrary, are fixed.  This is
reflected by the fact that the lapse and shift are fixed fields on ${\bf
R}\times\Sigma$, {\it i.e.}, not subject to variation in the action
principle.  Because of this, the field theory---in either the Lagrangian
or Hamiltonian formulation---is not ``generally covariant''. General
covariance can be introduced into the field theory by keeping the
spacetime metric fixed and adding the foliation itself to the space of
dependent variables to be varied in the action principle.  This is
conveniently done in the Hamiltonian formulation, where the new
dynamical variables are spacelike embeddings, which we shall denote by
$X^\alpha$. We must still introduce momenta $\Pi_\alpha$ conjugate to
the embeddings. To do this we need the unit normal $n^\alpha$ to
the hypersurface embedded by $X^\alpha$.  The unit normal is defined by
$$
 g_{\alpha\beta}(X)n^\alpha X^\beta_{,i}=0\qquad{\rm and}\qquad
g_{\alpha\beta}(X)n^\alpha n^\beta = -1,\tag 274
$$
 where
$g_{\alpha\beta}(X)$ is the metric on $\bf M$ restricted to the embedding
$X^\alpha$.  The unit normal is a fixed local function of the embedding
and its first spatial derivatives.  The foliation is a one-parameter
family of spacelike embeddings, which we shall denote by $X^\alpha(t)$.  It
is straightforward to show that
$$
 \dot X^\alpha(t) = N^\perp n^\alpha +
N^iX^\alpha_{,i}.\tag 275
$$
 The action \(271) can thus be written as
$$
I[\varphi^A,\pi_A]=\int_{{\bf R}\times\Sigma}\left(\pi_A\dot\varphi^A-\dot
X^\alpha h_\alpha\right),\tag 276
$$
 where
$$
 h_\alpha=-n_\alpha h_\perp
+X^i_\alpha h_i,\tag 277
$$
 and we have introduced fields $X_\alpha^i$,
which are fixed local functions of the embeddings and their first
spatial derivatives defined by
$$
 X_\alpha^i
X^\alpha_{,j}=\delta^i_j\qquad{\rm and}\qquad n^\alpha X_\alpha^i=0. \tag 278
$$
{}From this form of the action we see that the momenta conjugate to the
embeddings are given by
$$
 \Pi_\alpha=-h_\alpha;\tag 279
$$
 these
definitions represent constraints
$$
H_\alpha:=\Pi_\alpha+h_\alpha\approx0.\tag 280
$$
We can take the constraints
\(280) into account with Lagrange multipliers $N^\alpha$ and obtain the
final form for the Hamiltonian action describing the parametrized field
theory:
$$
 I[N^\alpha,X^\alpha,\Pi_\alpha,\varphi^A,\pi_A]=\int_{{\bf
R}\times\Sigma}\left(\pi_A\dot\varphi^A+\Pi_\alpha\dot X^\alpha
-N^\alpha H_\alpha\right).\tag 281
$$
 The extrema of this action,
obtained by varying it with respect to its arguments, are defined by a
system of equations equivalent to those obtained by extremising \(276).  Note
that the Hamiltonian can
be expressed as
$$
 {\bf H} := \int_\Sigma N^\alpha H_\alpha =
\int_\Sigma\left(N^\perp H_\perp +N^i H_i\right),\tag 282
$$
 where
$$
\eqalignno{ H_\perp&=n^\alpha \Pi_\alpha + h_\perp,&(283)\cr
H_i&=X^\alpha_{,i}\Pi_\alpha + h_i,&(284)}
$$
 and we can equally well
vary $N^\perp$ and $N^i$ instead of $N^\alpha$ in \(281).

The resulting formalism, in which the embeddings and their
conjugate momenta are adjoined to the phase space (at the expense of the
constraints \(280)) is the ``parametrized formalism'' for the
harmonic map field theory.  The central feature of the parametrized formalism
is that it provides a generally covariant formulation of any field theory.
This
is what makes possible the identification of the gravitational
models, which are
generally covariant field theories on $M$, with the theory of harmonic maps on
a fixed background spacetime.

Now we are ready to make contact with the Hamiltonian formulation of
cylindrically symmetric spacetimes.  This is accomplished in 3 steps.
\bigskip \line{\it 1. Fix the spacetime and target space.\hfill}

We fix the spacetime to be ${\bf M}={\bf R}^2\times{\bf S}^1$ equipped with
a flat metric, which is defined in
polar coordinates by the line element:
$$
 g_{\alpha\beta}dx^\alpha
dx^\beta = -dT^2 + dR^2 + R^2 d\Phi^2.\tag 285
$$
Here,
$T\in(-\infty,\infty)$, $R\in(0,\infty)$ and $\Phi\in(0,2\pi)$.  The harmonic
maps are
defined to take values in ${\bf L}={\bf R^2}$ with the metric $\sigma_{AB}$
chosen to be of constant negative curvature.  In coordinates
$\psi\in(-\infty,\infty)$ and $\phi\in(-\infty,\infty)$ the metric on
$\bf L$ is determined by the line element
$$
\sigma_{AB}d\varphi^Ad\varphi^B=d\psi^2 + e^{-2\psi}d\phi^2.\tag 286
$$
The scalar curvature of this metric is $-2$.

\bigskip \line{\it 2.  Impose azimuthal symmetry.\hfill}

Of course $\partial\hfill\over\partial\Phi$ is a Killing vector field of
the metric $g_{\alpha\beta}$.  We now demand that the fields $\varphi^A$ be
invariant along the flow generated by this Killing vector field, {\it
i.e.},
$$
 {\cal L}_{\partial\hfill\over\partial\Phi}\varphi^A=0.\tag 287
$$
 In the coordinates $(T,R,\Phi)$ on ${\bf M}$ this means we assume the
fields are independent of $\Phi$:
$$
 \psi=\psi(T,R)\qquad{\rm and}\qquad
\phi=\phi(T,R).\tag 288
$$
 In the parametrized formalism for the field
theory we also assume that the foliation is compatible with the
azimuthal symmetry in the sense that we only consider spacelike surfaces
to which $\partial\hfill\over\partial\Phi$ is everywhere tangent.  The
embeddings can be registered in the coordinates $(T,R,\Phi)$; that is,
we have
$$
 X^\alpha(r,\Phi)=(T(r),R(r),\Phi),\tag 289
$$
 where
$r\in(0,\infty)$. Thus, to specify an embedding we must specify two
functions of one variable, $T(r)$ and $R(r)$.  On the symmetry-compatible
foliation the shift vector takes the form
$$
 N^i=(N^r,0).\tag 290
$$

We can now formulate the parametrized harmonic map field theory on the
2-dimensional space of orbits $M={\bf R}\times {\bf R}^+$ of the Killing
vector.
By working in polar coordinates on ${\bf M}$, this amounts to simply ignoring
the $\Phi$ coordinate.  The
Hamiltonian form of the action can be obtained by substituting the
choices made in the first two steps into the action \(281) and
integrating out the angular coordinate.  We obtain
$$
 \eqalign{
I[N^\perp,N^r,T,R,&\Pi_T,\Pi_R,\psi,\phi,\pi_\psi,\pi_\phi]\cr
&=\int_{t_1}^{t_2}dt\,\int_0^\infty dr\,\left(\Pi_T\dot T +\Pi_R\dot R +
\pi_\psi\dot\psi + \pi_\phi\dot\phi -N^\perp H_\perp-N^rH_r\right),}\tag
291
$$
 where
$$
 \eqalignno{
H_\perp&={1\over\sqrt{R^{\prime2}-T^{\prime2}}} \Bigg[\Pi_TR^\prime
+
\Pi_RT^\prime +{1\over2}\left(R\psi^{\prime2}+R^{-1}\pi_\psi^2\right)\cr
&\qquad\qquad\qquad+{1\over2}\left(Re^{-2\psi}\phi^{\prime2}+R^{-1}e^{2\psi}
\pi_\phi^2\right)\Bigg],
&(292)\cr
H_r&=\Pi_TT^\prime +\Pi_RR^\prime + \pi_\psi\psi^\prime
+\pi_\phi\phi^\prime.&(293)}
$$
 Variation of this action with respect to
its arguments yields field equations equivalent to those obtained {}from
\(281) in the special case of azimuthal symmetry.  An equivalent
variational principle is obtained by defining a lapse density
$$
N:={1\over\sqrt{R^{\prime2}-T^{\prime2}}}N^\perp,\tag 294
$$
 and
weight-two super-Hamiltonian
$$
 H=\Pi_TR^\prime + \Pi_RT^\prime
+{1\over2}\left(R\psi^{\prime2}+R^{-1}\pi_\psi^2\right)
+{1\over2}\left(Re^{-2\psi}\phi^{\prime2}+R^{-1}e^{2\psi}\pi_\phi^2\right).\tag
295
$$
In terms of these quantities we have
$$
 \eqalign{
\bar I[N,N^r,T,R,&\Pi_T,\Pi_R,\psi,\phi,\pi_\psi,\pi_\phi]\cr
&=\int_{t_1}^{t_2}dt\,\int_0^\infty dr\,\left(\Pi_T\dot T +\Pi_R\dot R +
\pi_\psi\dot\psi + \pi_\phi\dot\phi -N H-N^rH_r\right),}\tag 296
$$
 and
variation of this action with respect to its argument yields field
equations equivalent to those obtained {}from \(291).  Note the
close similarity
between the super-Hamiltonian \(295) for the harmonic map theory and the
super-Hamiltonian \(268) for cylindrically symmetric spacetimes.

\bigskip \line{\it 3.  Canonical transformation.\hfill}

In the last step, we perform a canonical transformation that
interchanges the roles of $\phi$ and $\pi_\phi$ and puts the
super-Hamiltonian and super-momentum of the parametrized field theory of
harmonic maps into the form \(268) and \(269).
This transformation is given by
$$
 \eqalignno{ \psitilde(r)&=\int_\infty^r
d\bar r\,\pi_\phi(\bar r)&(303)\cr \pi_\psitilde&=\phi^\prime.&(304)}
$$
The inverse transformation is
$$
 \eqalignno{ \phi(r)&=\int_0^r d\bar
r\,\pi_\psitilde(\bar r)&(301)\cr \pi_\phi&=\psitilde^\prime.&(302)}
$$
With the boundary conditions given in the Appendix, this is a canonical
transformation provided we impose the boundary condition $\phi(r=0)=0$.
It follows immediately {}from \(301) and \(302) that, in the new
variables, the super-Hamiltonian and super-momentum for the
parametrized
harmonic map field theory are precisely \(268) and \(269).

As indicated by the equivalence to a parametrized field theory,
the constraints
\(268) and \(269),
$$
 H\approx0\approx H_r,\tag 297
$$
 can be solved for the
momenta conjugate to the embeddings.  Indeed, it is straightforward to
show that these constraints are equivalent to
$$
 H_a:=\Pi_a +
h_a\approx 0,\tag 298
$$
 where $\Pi_a=(\Pi_T,\Pi_R)$, and $h_a=(h_T,h_R)$ is defined as in \(277)
with
$$
n_a=(n_T,n_R)=\left(-{R^\prime\over\sqrt{R^{\prime2}-T^{\prime2}}}
,{T^\prime\over\sqrt{R^{\prime2}-T^{\prime2}}}\right),\tag 299
$$
$$
X^r_a=(X^r_T,X^r_R)=\left(-{T^\prime\over
R^{\prime2}-T^{\prime2}},{R^\prime\over R^{\prime2}-T^{\prime2}}\right).
\tag 29901
$$
Explicitly, the constraints \(298) take the form
$$
\eqalign{
\Pi_T
&+{1\over{R^{\prime2}-T^{\prime2}}}\Bigg[R^\prime\left({1\over2}
\left(R\psi^{\prime2}+R^{-1}\pi_\psi^2\right)
+{1\over2}\left(Re^{-2\psi}\pi_\psit^2+R^{-1}e^{2\psi}\psit^{\prime2}\right)
\right)\cr
&-T^\prime\left(\pi_\psi\psi^\prime
+\pi_\psit\psit^\prime\right)\Bigg]\approx 0,}\tag3111
$$
and
$$
\eqalign{
\Pi_R
&-{1\over{R^{\prime2}-T^{\prime2}}}\Bigg[T^\prime\left({1\over2}
\left(R\psi^{\prime2}+R^{-1}\pi_\psi^2\right)
+{1\over2}\left(Re^{-2\psi}\pi_\psit^2+R^{-1}e^{2\psi}\psit^{\prime2}\right)
\right)\cr
&-R^\prime\left(\pi_\psi\psi^\prime
+\pi_\psit\psit^\prime\right)\Bigg]\approx 0.}\tag3002
$$
A standard argument (see, {\it e.g.}, \refto{CGT1991}) establishes that the
constraints \(3111) and \(3002)
have an Abelian Poisson bracket algebra:
$$
\{H_a(r),H_b(\bar r)\}=0.\tag 3000
$$

To summarize, the Hamiltonian structure of cylindrically symmetric
gravitational fields is mathematically identical to a parametrized field
theory of azimuthally symmetric harmonic maps {}from a 3-dimensional
flat spacetime to a 2-dimensional manifold equipped with a metric of constant
negative curvature. It is important to note that all of these results
are fully gauge invariant in the sense that no coordinate conditions are
needed to be imposed on the gravitational theory.  In any of the forms
that we have presented the Hamiltonian formulation, the field theory retains
the full 2-dimensional diffeomorphism invariance compatible with the
imposition of cylindrical symmetry.

\bigskip \line{\bf 3b.  Closed universes: Toroidal Symmetry.\hfill}

We now repeat the analysis of \S3a under the assumption of toroidal
symmetry.  The procedure is very similar to that used in the cylindrical
symmetry case; the key difference is the way in which the missing
degree of freedom is introduced.

Again, the strategy is to use the conformal coordinates $\tau$ and $X$
to define embeddings of a circle into $M={\bf R}^+\times{\bf S}^1$.  The
variable $\tau$ already appears as a canonical coordinate on the phase
space, but we must still express $X$ as a function on phase space.  This
can be achieved starting {}from \(237).  Choosing the spacelike circle on
which the integral is performed to be a $t=const$ slice, we have
that
$$
 X^\prime={1\over\pi_0}\pi_\gamma,\tag 305
$$
 where
$$
\pi_0:={1\over2\pi}\int_0^{2\pi} dx\, \pi_\gamma(x).\tag 306
$$
Note that, because of \(249),
$\pi_\gamma$ is negative-definite, so the denominator never
vanishes in \(305) and $X(x)$ is monotonic.

Before completing the phase space definition of $\tau$ and $X$ into a
canonical transformation we have to contend with the fact that \(305)
does
not define $X$ as a function on the gravitational phase space $\Gamma$.
The reason is the same as in the cylindrically symmetric case: there is
an integration constant left unspecified in \(305), which is not fixed by the
gravitational phase space data.  In \S3a the
integration
constant represented the asymptotic location of a spacelike slice, here
it represents the relation between the origin of the coordinate $x$ on
${\bf S}^1$ and the origin of the conformal coordinate $X$. This information
is coordinate-dependent and not included in the gravitational phase
space.  As before, we remedy this situation by adding a new degree of
freedom $q$ to the phase space.  Unlike the cylindrically symmetric
case, this new degree of freedom has no role to play in gravitational
dynamics, {\it i.e.}, it is pure gauge.  We therefore introduce a momentum
$p$ conjugate to $q$ and adjoin a new constraint,
$$
p\approx0,\tag 307
$$
to
the Hamiltonian formulation.  We denote by $\Gamma^\star$ the phase
space extended by the variables $q$ and $p$.  Our strategy is to
formulate the gravitational system as a parametrized field theory on
$\Gamma^\star$ and then reduce the system by the constraint \(307).
When
we reduce $\Gamma^\star$ by the constraint \(307) we arrive at the
original
gravitational phase space $\Gamma$ and the dynamics thereon.

We are now ready to extract embedding variables from the gravitational system
formulated on $\Gamma^\star$.  Let $\mu(x)$ be a
prescribed measure on the circle, {\it i.e.}, $\mu$ is a positive
density of weight-one on the circle, and is normalized via
$$
\int_0^{2\pi}dx\,\mu(x) = 1.\tag 308
$$
We define a transformation
$$
 (\tau,\gamma,
\psi,\tilde\psi,q, \pi_\tau,\pi_\gamma, \pi_\psi,\pi_{\tilde\psi},
p) \longleftrightarrow (T,X,\phi,\tilde\phi,{\cal Q},
\Pi_T,\Pi_X,\pi_\phi,\pi_{\tilde\phi}, {\cal P}),\tag 309
$$
by
$$
 \eqalignno{ T&=
-{1\over\pi_0}\tau,&(311)\cr
X(x)&=q +
\int_0^{2\pi}dx^{\prime\prime}\,\mu(x^{\prime\prime})\int_{x^{\prime
\prime}}^x dx^\prime\,{1\over\pi_0}\pi_\gamma(x^\prime),&(312)\cr
\Pi_T&=-\pi_0\left(\pi_\tau + \left[\ln\left({ \pi_\gamma-
\tau^\prime\over\tau^\prime+\pi_\gamma}\right)\right]^\prime\right),
&(313)\cr
\Pi_X&=p\mu + \pi_0\left(\gamma^\prime - \left[\ln(
\pi_\gamma^2-\tau^{\prime2})\right]^\prime\right),&(314)}
$$
 and
$$
\eqalignno{ \phi&=\sqrt{-\pi_0}\psi,&(315)\cr
\tilde\phi&={1\over\sqrt{-\pi_0}}\psit,&(316)\cr
\pi_\phi&={1\over\sqrt{-\pi_0}}\pi_\psi,&(317)\cr
\pi_{\tilde\phi}&=\sqrt{-\pi_0}\pi_{\tilde\psi},&(318)\cr
{\cal Q}&={1\over\pi_0}\int_0^{2\pi}dx\,\Bigg\{\left(\gamma-
\ln(\pi_\gamma^2-\tau^{\prime2})\right)\pi_\gamma
-\left(\pi_\tau+\left[\ln\left({ \pi_\gamma-
\tau^\prime\over\tau^\prime+\pi_\gamma}\right)\right]^\prime\right)\tau\cr
&\qquad\qquad\qquad+{1\over2}\pi_\psi\psi -
{1\over2}\pi_{\tilde\psi}\tilde\psi\Bigg\},&(319)\cr
{\cal P}&=\pi_0.&(320)}
$$
 Here
we have made some convenient rescalings. In particular, the scaling by
${1\over\pi_0}$ used in the definition of $X$ guarantees that
$$
X(2\pi)-X(0)=2\pi.\tag 321
$$
We mention that the requirement that
$D_a\tau$ is timelike, given by \(249), is equivalent to
$$
X^{\prime}>|T^{\prime}|.\tag 322
$$
The inequalities \(249) and \(322) guarantee that
the above transformation is well-defined and that the slices embedded by
$(T(x),X(x))$ are spacelike. Let us also note that the constraint \(307) is, in
the
new
variables, the constraint
$$
 \int_0^{2\pi}dx\,\Pi_X(x)\approx0.\tag 323
$$
It is
not too hard to show, {\it e.g.}, by expanding in Fourier series, that
the above transformation is a bijection {}from $\Gamma^\star$ to the
phase space $\Upsilon$ for a field theory of
$(\phi,\tilde\phi,{\cal
Q},\pi_\phi,\pi_\phit,{\cal P})$ in the parametrized
formalism.  Our notation here is that $\Upsilon$ is the product of the
cotangent
bundle over the space of embeddings of a circle into $M$, and
the phase space of the canonical variables $(\phi,\tilde\phi,{\cal
Q})$ and $(\pi_\phi,\pi_\phit,{\cal P})$.
By computing Poisson brackets, or by
computing the symplectic form in the new variables, it is
straightforward to verify that the transformation is canonical, \ie
identifies the respective symplectic structures, and that the variables
$
 (T,X,\phi,\tilde\phi,{\cal Q})$ and
$(\Pi_T,\Pi_X,\pi_\phi,\pi_{\tilde\phi}, {\cal P})$
are canonical coordinates and momenta.

Modulo the constraint \(323), the weight-2 super-Hamiltonian and
super-momentum take the following form when expressed in terms of the
new canonical variables:
$$
H:=\Pi_TX^\prime + \Pi_X
T^\prime+ {1\over2}\left(T\phi^{\prime2}+{1\over T}\pi_{\phi}^2\right)
+{1\over2}\left(T e^{-{2\over\sqrt{-\cal P}}\phi}\pi_{\tilde\phi}^2
+{1\over T}e^{{2\over\sqrt{-\cal P}}\phi}\tilde\phi^{\prime2}\right)
\approx0,
\tag324
$$
and
$$
H_x:=\Pi_T T^\prime + \Pi_X X^\prime +\pi_\phi\phi^\prime
+\pi_{\tilde\phi}\tilde\phi^\prime\approx0.\tag325
$$
The constraints of the
theory are \(323)--\(325); they are ``first-class''. In particular, the
Poisson brackets of the constraint \(323) with the super-Hamiltonian and
super-momentum vanish because \(323) generates constant shifts of $X$
with
respect to which $H$ and $H_x$ are invariant.

Let us make one further remark about the canonical transformation \(309).
First,
we emphasize that the identification of $\Gamma^\star$ with
$\Upsilon$ depends on the non-gravitational data $(q,p,\mu)$.  However,
upon passing to $\Gamma$ using the constraint \(323) this dependence
necessarily
disappears. To see this, recall that the reduction from $\Gamma^\star$ to
$\Gamma$ is obtained by (i) restricting the phase space to the constraint
surface defined by \(323), and (ii) identifying points on the constraint
surface
which lie on an orbit of the canonical transformations generated by
$\int_{{\ss {\bf S}^1}}\Pi_X$, that is, embeddings $X$ and $X+const$
are identified. The
dependence of $\Pi_X$ on $p$ and $\mu$ is eliminated upon passing to the
constraint surface defined by \(323). Restricting to the constraint surface
of \(323), we must identify $X$ and $X+const$.  For a fixed choice of
$\mu$ this eliminates the dependence of the phase space on $q$.  Now suppose we
use a different measure
$\tilde\mu$ to define a new embedding variable
$\tilde X(x)$.  Because $\tilde
X^\prime=\pi_\gamma/\pi_0$ we have that $(\tilde X-X)^\prime=0$, that
is, $\tilde X=X + const$, which is precisely the transformation generated by
$\int_{{\ss {\bf S}^1}}\Pi_X$, so on $\Gamma$ the phase space is independent
of the choice
of measure $\mu$.  Put differently, changing $\mu$ is equivalent to holding
$\mu$ fixed and changing $q$, which has no effect on points of $\Gamma$.
Thus the non-gravitational data $(q,p,\mu)$ are eliminated upon passing from
$\Gamma^\star$
to the original phase space $\Gamma$.

Once again we can interpret the resulting formalism in terms of a
parametrized field theory of harmonic maps.  However, it must be kept in
mind that there is a ``point particle'' degree of freedom
represented by $\cal Q$ and $\cal P$ along with an extra constraint
\(323).
Because $\cal Q$ is cyclic in the Hamiltonian, we can reduce the phase
space $\Gamma^\star$ by the integral of motion $\cal P$ to a phase
space $\tilde\Gamma^\star$.  On $\tilde\Gamma^\star$ we view $\cal
P$
as a parameter, which will appear in the metric for the spacetime upon which
the harmonic maps are defined.  To express the gravitational theory
formulated on $\tilde\Gamma^\star$ as a parametrized harmonic map
field
theory we repeat the 3 steps of \S3a.

\bigskip \line{\it 1. Fix the
spacetime and target space.\hfill}

The spacetime is now taken to be ${\bf M}={\bf R}^+\times {\bf T}^2$ with
metric given by the line element
$$
 ds^2=l^2\left(-dT^2 + dX^2 + T^2 dY^2\right),\tag 327
$$
where $X$ and $Y$ are coordinates on ${\bf T}^2$, $T>0$, and $l$ is a positive
constant. Using an analogous
construction to that found in
\S2b, this metric can be viewed as a flat metric on a compactification
of the chronological future $I^+$ of the origin of 3-dimensional Minkowski
spacetime
${\bf M}^3$.  Let us spell this out in detail.  Let $\hat t$, $\hat x$,
$\hat y$
denote inertial coordinates on ${\bf M}^3$.  On $I^+$ make the identification
$$
\hat x\sim\hat x + 2\pi l,\qquad
\tanh^{-1}({\hat y\over\hat t})\sim\tanh^{-1}({\hat y\over\hat t})
+2\pi ,\tag 37900
$$
and define the resulting manifold by $\tilde I$.
The Minkowski metric on ${\bf M}^3$ projects to a flat metric on $\tilde I$.
Define coordinates $(T,X,Y)$ on ${\bf R}^+\times {\bf T}^2$, where $T\in(0,
\infty)$, $X\in(-\infty,\infty)$, and $Y\in(-\infty,\infty)$, with $X\sim X+
2\pi$ and $Y\sim Y+2\pi$. The diffeomorphism $\phi\colon{\bf R}^+
\times{\bf T}^2\to\tilde I$, defined by
$$
\eqalign{
\hat t&= l\, T \cosh Y,\cr
\hat x&= l\, X,\cr
\hat y&=l\, T\sinh Y,}\tag 37901
$$
identifies the  metric \(327) on ${\bf R}^+\times {\bf T}^2$ with the Minkowski
metric on $\tilde I$.

As in the cylindrically symmetric case, the
harmonic maps $\varphi^A:=(\alpha,\beta)$
take values in ${\bf L}={\bf R}^2$, which is equipped with a
metric of constant negative curvature given by the line element
$$
\sigma_{AB}d\varphi^Ad\varphi^B=d\alpha^2+e^{-2\alpha}d\beta^2.\tag 328
$$

\bigskip \line{\it 2. Impose azimuthal symmetry.\hfill}

The vector field ${\partial\hfill\over\partial Y}$ is a Killing vector
field for the metric on $\bf M$.  We now demand that all the fields of the
parametrized harmonic map field theory are likewise invariant under the
1-parameter family of isometries generated by
${\partial\hfill\over\partial Y}$.  Thus we can formulate the theory on
the space of orbits, $M={\bf R}^+\times{\bf S}^1$, of the Killing vector field.
On the space of orbits of ${\partial\hfill\over\partial Y}$, the
Hamiltonian form of the action is given by
$$
 \eqalign{
\bar I[N,N^x,T,X,&\pi_T,\pi_X,\alpha,\beta,\pi_\alpha,\pi_\beta]\cr
&=\int_{t_1}^{t_2}dt\,\int_0^{2\pi} dx\,\left(\pi_T\dot T +\pi_X\dot X +
\pi_\alpha\dot\alpha + \pi_\beta\dot\beta -N H-N^xH_x\right),}\tag 329
$$
where
$$
H=\pi_TX^\prime + \pi_XT^\prime
+{1\over2}\left((l\,T)^{-1}\pi_\alpha^2+l\,T\alpha^{\prime2}\right)
+{1\over2}\left((l\,T)^{-1}e^{2\alpha}
\pi_\beta^2+l\,Te^{-2\alpha}\beta^{\prime2}
\right),\tag 330
$$
and
$$
 H_x=\pi_TT^\prime +\pi_XX^\prime +
\pi_\alpha\alpha^\prime +\pi_\beta\beta^\prime.\tag 331
$$
Note the similarity between the super-Hamiltonian \(330) for the harmonic map
theory and the super-Hamiltonian \(324) for toroidally symmetric spacetimes.

\bigskip \line{\it 3.  Canonical transformation.\hfill}

In this last step we make a canonical transformation which puts the
super-Hamiltonian and super-momentum \(330) and \(331) of the
parametrized field
theory into the form \(324) and \(325) found in the gravitational
theory.  It is
possible to adapt the transformation \(303)--\(304) used for the
cylindrically
symmetric case, but this leads to additional---and
unnecessary---constraints on the harmonic maps.  In the toroidal
symmetry case the following canonical transformation accomplishes our goal
\refto{ctnote}.
$$
\eqalignno{
\alpha&=-l^{-{1\over2}}\phi + \ln(l\,T),&(336)\cr
\beta&=l^{1\over2}\phit,&(3390)\cr
\pi_\alpha&=-l^{1\over2}\pi_\phi + l\,X^\prime,&(337)\cr
\pi_\beta&=l^{-{1\over2}}\pi_\phit,&(3370)\cr
\pi_T&=\Pi_T + l^{1\over2}T^{-1}\pi_\phi - l\,(2T)^{-1}X^\prime,&(338)\cr
\pi_X&=\Pi_X+l^{1\over2}\phi^\prime-l\,(2T)^{-1}T^\prime.&(3391)}
$$
Note that the embedding coordinates $(T,X)$ retain
their original meaning, and the embedding momenta, while redefined, are
still scalar densities of weight-one on the circle.   By using the
transformation \(336)--\(3391) it follows that the super-Hamiltonian and
super-momentum \(330) and \(331) of the
parametrized field
theory become those  found in the gravitational
theory formulated on $\tilde\Gamma^\star$ (\(324) and \(325)), provided we make
the identfication
$$
l=-{\cal P}.\tag 33100
$$

As in the cylindrically
symmetric case, we now know that the constraints \(324) and \(325),
$$
 H\approx0\approx H_x,\tag 332
$$
can be
reexpressed as
$$
 H_a:=\Pi_a + h_a\approx 0,\tag 333
$$
where
$\Pi_a=(\Pi_T,\Pi_X)$, and $h_a=(h_T,h_X)$ is defined as in \(277)
with
$$
n_a=(n_T,n_X)=\left(-{X^\prime\over\sqrt{X^{\prime2}-T^{\prime2}}}
,{T^\prime\over\sqrt{X^{\prime2}-T^{\prime2}}}\right),\tag 334
$$
$$
X^x_a=(X^x_T,X^x_X)=\left(-{T^\prime\over
X^{\prime2}-T^{\prime2}},{X^\prime\over X^{\prime2}-T^{\prime2}}\right).
\tag 335
$$
Explicitly, the constraints \(333) take the form
$$
\eqalign{
\Pi_T
&+{1\over{X^{\prime2}-T^{\prime2}}}\Bigg[X^\prime\left({1\over2}
\left(T\phi^{\prime2}+T^{-1}\pi_\phi^2\right)
+{1\over2}\left(Te^{-2\phi/\sqrt{-\cal P}}\pi_\phit^2
+T^{-1}e^{2\phi/\sqrt{-\cal P}}\phit^{\prime2}\right)
\right)\cr
&-T^\prime\left(\pi_\phi\phi^\prime
+\pi_\phit\phit^\prime\right)\Bigg]\approx 0,}\tag3222
$$
and
$$
\eqalign{
\Pi_X
&-{1\over{X^{\prime2}-T^{\prime2}}}\Bigg[T^\prime\left({1\over2}
\left(T\phi^{\prime2}+T^{-1}\pi_\phi^2\right)
+{1\over2}\left(Te^{-2\phi/\sqrt{-\cal P}}\pi_\phit^2
+T^{-1}e^{2\phi/\sqrt{-\cal P}}\phit^{\prime2}\right)
\right)\cr
&-X^\prime\left(\pi_\phi\phi^\prime
+\pi_\phit\phit^\prime\right)\Bigg]\approx 0.}\tag3223
$$
As before, the constraints \(3222) and \(3223) have an Abelian Poisson bracket
algebra:
$$
\{H_a(x),H_b(\bar x)\}=0.\tag 3001
$$

To summarize, the Hamiltonian formulation on $\tilde\Gamma^\star$ of
toroidally symmetric gravitational fields is equivalent to that of a
parametrized field theory of harmonic maps {}from a flat 3-dimensional
spacetime to a
2-dimensional manifold of constant negative curvature.  Of course, to
recover the original gravitational field theory on $\Gamma$ the degrees
of freedom $({\cal Q}, {\cal P})$ must be reinstated but, more
importantly, the constraint \(323) must be imposed on the harmonic map
field
theory.   This constraint sets to zero the homogeneous mode of the
embedding momentum $\Pi_X$, and, by virtue of \(3223), is equivalent to
the
constraint
$$
 \int_0^{2\pi}dx\,h_X(x)\approx0.\tag 340
$$
This constraint can be
viewed as setting to zero the total momentum of the fields $\phi$ and $\phit$
(or $\alpha$
and $\beta$). Because the constraint
\(340)
is first-class, we are obliged to identify $X$ and $X+const$.  The equivalence
class of embeddings thus obtained is
independent of the value of the non-gravitational variables $(q,p,\mu)$.  On
the other hand,
the extra constraint means that, strictly speaking, the Hamiltonian formulation
of toroidally symmetric
spacetimes is not quite identical to a parametrized field theory of harmonic
maps.  However, as we shall see, the correspondence is sufficiently
close to allow a formal construction of the quantum theory.

\taghead{4.}
\bigskip
\line{\bf 4. Implications for quantum theory.\hfill}

Issues which arise in canonical quantum gravity, the problems of time in
particular \refto{Kuchar1992}, can be investigated in the
models we have been studying. In the last section we exhibited
canonical transformations that identify certain phase spaces for these models
with phase spaces for harmonic maps from flat 3-dimensional
spacetimes into a target space of constant negative curvature.  This
mathematical identification is fully gauge invariant, {\it i.e.}, preserves the
2-dimensional diffeomorphism symmetry exhibited by the two
Killing vector models.
It should be emphasized, however, that the extraction of the embedding
variables was not
without its complications.  We encountered a ``global problem of time'' which
was resolved by adding a finite number of non-gravitational degrees of freedom
to the usual ADM phase space. Given the extended phase spaces, each event in
the
effective two-dimensional spacetime is uniquely labeled by the values of the
canonical variables $X^a$ on a spacelike slice at the point where the spacelike
slice passes through that event.  The identification of spacetime points
provided by $X^a$ is independent of the choice of slice and the ``spacetime
problem'' \refto{Kuchar1992} is avoided.  Given these results, the obvious
strategy for
quantization is to
view the dynamics of the models as a generally covariant ({\it i.e.},
parametrized) formulation of non-linear fields on a fixed background.  The
quantum theory of fields on a fixed spacetime is quite well-studied and, at
least superficially, presents no overwhelming conceptual
difficulties---although
the quantum theory of interacting fields is always technically challenging. A
simplifying feature of these models is that, while they can be viewed as field
theories on a 3-dimensional flat spacetime, the Killing vector structure is
such
that the final result is in each case a 2-dimensional field theory, and in two
dimensions quantum field theory is typically more manageable than in higher
dimensions.   Our purpose in this section is to discuss certain broad features
of the quantization of these models based upon the classical structures
elucidated
in \S3.  We hope to return to a more detailed examination of
the resulting quantum theories in future work.

In each of the models we have studied the resulting Hamiltonian structure
involves embedding variables $X^a$ and their conjugate momenta $\Pi_a$,
along with dynamical variables $(q^{\bf\ss A},p_{\bf\ss A})$ which are
$(\psi, \psit,\pi_\psi,\pi_{\psit})$ in the
cylindrical symmetry case, or $({\cal Q},\phi,\tilde\phi,{\cal
P},\pi_{\ss \phi},\pi_{\ss
\tilde\phi})$ in the toroidal symmetry case.  The constraints include
diffeomorphism constraints \(298), \(333) and, in the case of toroidal
symmetry,
an
additional constraint \(323). Ignoring its origins, the parametrized field
theory
defined by these
constraints is, at least formally, relatively straightforward to ``quantize''.
Note however that in each of the models the effective
2-dimensional field theory has a complicating feature not usually found in more
familiar 2-dimensional field theories.  Namely, the super-Hamiltonian is an
explicit function of $R(r)$ in the cylindrical symmetry case, and $T(x)$ in the
toroidal symmetry case.  This will modify the quantum theories relative to what
we might find, {\it e.g.}, for a conformal field theory.
In any case, we promote the canonical variables $(q^{\bf\ss A},p_{\bf\ss A})$
to
operators on states $|\Psi>$. One way to do this would be, using the Heisenberg
picture, to quantize
the variables $(q^{\bf\ss A},p_{\bf\ss A})$ on the priveleged foliation $T=t$
and
$R=r$ or $X=x$.  The states are embedding-independent
in the Heisenberg picture; the operators are evolved by the
constraint operators \refto{Kuchar1989b}. Of course it may be
necessary to use perturbation theory
to define the operators and states.
Formally, observables are self-adjoint operators representing classical
functionals of $(q^{\bf\ss A},p_{\bf\ss A})$.  Note that in the toroidal
symmetry
model, observables must commute with the constraint \(323).

In the Schr\"odinger picture, dynamical evolution corresponds to considering
states $|\Psi,X^a>$ which are parametrized by the embeddings.  The states are
evolved from one embedding to the
next by the energy-momentum current $h_a$.
More
precisely, the states $|\Psi,X^a>$ are defined as solutions to the functional
Schr\"odinger equation
$$
i{\delta\over\delta X^a}|\Psi,X^a>
=h_a|\Psi,X^a>.\tag 41
$$
This equation can be considered an implementation of the demand that
``physical states are annhilated by constraints'': $H_a|\Psi,X^a >=0$.  In the
toroidal symmetry case we
must also impose the quantum version of \(323):
$$
\left(\int_{0}^{2\pi}dx\,{\delta\over\delta X(x)}\right)
|\Psi,X^a> = 0.\tag 42
$$
This requirement is equivalent to
$$
|\Psi,T,X+ \ const>=|\Psi,T,X>,\tag 43
$$
but, in light of \(41) is also equivalent to
$$
\left(\int_0^{2\pi}dx\ h_X(x)\right)|\Psi,X^a>=0.\tag 44
$$
Given an initial state $|\Psi_0, X^a_0>$ on an initial embedding
$X^a_0$, the functional
Schr\"odinger equation \(41) is solved (subject to the subsidiary
condition \(42)) and
the solution is matched to the initial data.  Note that \(42) can be
implemented by
imposing \(43) on the initial state $|\Psi_0, X^a_0>$, and then solving \(41).
If \(42) is
satisfied initially, it will be satisfied on any embedding provided the state
is evolved according to \(41).  The resulting embedding-dependent vector
$|\Psi,X^a>$
is to be interpreted as the state of the
system on the embedding $X^a(x)$ and one can go on to predict
outcomes of measurements of observables constructed from
the quantum fields (and
$X^a(x)$ if desired) on various hypersurfaces in the
usual way.

For this quantization strategy to be feasible, one must check
that the operator representatives of the currents $h_a$ are well defined
operator-valued distributions.  Furthermore, one must guarantee that the
infinite number of differential equations contained in \(41), \(42) are
mutually
consistent.  The integrability conditions for this are the quantum commutator
analogs of
the Abelian algebra \(3000) and \(3001), which guarantee that quantum dynamics
does not depend on the choice of observer. More precisely, the integrability
conditions for \(41) imply that the state $|\Psi_1,X^a_1 >$ obtained by
evolving
an initial state $|\Psi_0,X_0^a >$ along a specific foliation connecting the
slices
defined by $X_0^a$ and $X_1^a$\ does not depend on the choice of the
foliation. For these integrability conditions to be
satisfied, it
is essential that the commutators of the components of the energy-momentum
current
are, up to a factor of $i$, the same as their classical Poisson brackets.
It is
very likely that this will {\it not} happen, {\it i.e.}, anomalies (or
``Schwinger terms'') terms will
arise.  This is the ``functional evolution problem'' \refto{Kuchar1992}.  It
would be very
interesting  to compute these Schwinger
terms.  If they depend only on the embedding, and are finite, then one
can use techniques developed by Kucha\v r for free fields in two
dimensions, where the anomalous terms are finite and depend only on the
embeddings \refto{Kuchar1989b}.  If the anomalous terms are operator-valued, or
infinite, it is
not at all clear how to proceed.  Note that if we turn off one of the
polarizations of the
gravitational field by setting $\psit=0$, then the resulting field theory is
linear.  In this case it would seem that all operators can be defined by normal
ordering, and that the Schwinger terms are finite and depend only on the
embeddings, but to our knowledge no one has computed the Schwinger terms
for this interesting special case.  In many ways, the functional evolution
problem is the most important issue to address in studying the quantization of
the two Killing vector models,
for it is here that one verifies the compatibility of the
method of quantization with the principle of general covariance.

Given a successful resolution of the functional evolution problem, we should
relate the elements of the quantum parametrized theory
with geometrical elements of spacetime.  In particular, what
does it mean to ``know the state on the embedding $X^a(x)$''?
Let us treat each
of the models in turn.  To fix
the state $|\Psi, X^a>$ of a cylindrically symmetric spacetime, we imagine a
family of observers and the associated spacelike foliation of spacetime.  On a
given slice of the foliation the observers measure the gravitational variables
$(R,\pi_\gamma)$ and the
``laboratory variable'' $\tau_\infty$; this fixes the
embedding of the slice,
$X^a(r)=(T(r),R(r))$, via \(260).
The observers
also measure measure a complete set of
commuting observables built from the operator representatives of $(\psi,
\psit,\pi_\psi,\pi_\psit)$.   All together, these measurements fix the
quantum state on a spacelike hypersurface in terms of measurement of
geometrical
quantities and the reading of a clock at infinity.
In the toroidal symmetry case
the family of observers measure $\tau/\pi_0$ and
$\pi_\gamma/\pi_0$; this determines the embedding $X^a(x)=(T(x),X(x))$
via \(311) and \(312) up to
specification of the non-gravitational variables $q$ and
$\mu$. The observers also measure a complete set of commuting observables built
from
$({\cal Q},\phi,\tilde\phi,{\cal P},\pi_{\ss \phi},\pi_{\ss
\tilde\phi})$, which are defined in terms of the spacetime geometry in
\(315)--\(320).
Different choices of $q$ and
$\mu$ simply
redefine $X(x)$ by addition of a constant.   Because the states of a toroidally
symmetric spacetime must be
invariant under this transformation (see \(43)), the quantum state
$|\Psi,X^a>$ is
unambiguously determined by measurements of gravitational data only.

There is one remaining problem that should be discussed when
quantizing general relativity as a parametrized field theory.
Kucha\v r calls this problem the ``multiple choice
problem'' \refto{Kuchar1992}. The
quantization we have outlined apparently depends quite heavily on the way in
which the canonical variables $X^a$ are used to identify points of space and
instants of time.  To be sure, the embeddings we have constructed are, in some
sense, geometrically natural. But it is conceivable that other, equally valid,
embedding variables could be constructed. In this case it is not at all clear
that when using some other embedding variables the
resulting quantum theory will
coincide with the one we have outlined here. If the quantum theories based on
different choices of embedding variables do not coincide, {\it i.e.}, are not
physically equivalent, then we have an embarassment of riches: which quantum
theory describes the real world?  This issue can be examined in the context of
the parametrized formalism for a
relativistic particle moving in a curved spacetime, and it is found that the
multiple
choice problem can be quite severe \refto{Kuchar1992}.  In many ways it is the
multiple choice
problem which most deeply reflects the conflict
between general relativity and quantum mechanics. In order to
examine this problem in the models we have considered, new sets of embedding
variables are required. For example, one relatively simple way to obtain other
embedding variables is to take the embeddings $X^a$ we have constructed here
and perform the point transformation
$$
     \widetilde X^a:=F^a(X).\tag 45
$$
Assuming $F^a$ is smooth and admits a smooth inverse,
this transformation can be
easily completed to a canonical transformation.  These new embeddings can be
interpreted as follows.  Fix an embedding $X^a$ of a hypersurface $\Sigma$ in
spacetime.  Displace $\Sigma$ by letting the diffeomorphism $F^a$ act on it
pointwise.  The resulting hypersurface is embedded by $\widetilde X^a$.
Is it possible to
construct a quantum theory of the parametrized field theories representing the
classical two Killing vector models so that
transformations such as \(45) lead to physically equivalent theories?
Of course we
cannot answer this question here since we have only outlined the most basic
features to be expected in the putative quantum theory. Let us however suggest
that the multiple choice problem is closely allied with a familiar issue that
arises in quantization of a gauge theory using gauge fixing conditions.  There,
the question is whether or not the predictions of quantum field theories based
on different gauge fixing conditions agree.  The connection of this issue with
the multiple choice problem stems from the fact that, at least in the models
considered here, the classical gravitational theory can be obtained by (i)
choosing the gauge $T=t$ and $R=r$ or $X=x$, and (ii) ``parametrizing''
the resulting field theory to regain general covariance.  Likewise, we quantize
the fields $(q^{\ss\bf A},p_{\ss\bf A})$ on the foliation $T=t$ and $R=r$ or
$X=x$,
and then reinstate general covariance by constructing states parametrized by
the
embeddings $X^a$ which satisfy \(41).  Whether general covariance is in fact
realized in the quantum theory is determined by the outcome of the problem of
functional evolution.  Let us assume that this problem can be solved.  The
remaining question is how the quantum theory depends on the initial choice of
gauge.  In gauge theories, this issue is fruitfully analyzed using
BRST methods \refto{Henneaux1992} which, at least formally, guarantee the
independence of the
predictions of the quantum theory from the underlying choice of gauge.  It
remains to be seen whether the multiple choice problem in generally covariant
theories can be resolved in a
similar way.
\bigskip\bigskip
\line{\bf Acknowledgments.\hfill}
The authors would like to thank Karel Kucha\v r and
Madhavan Varadarajan for many helpful discussions.
J.D.R.~was supported by grant NSF-PHY-92-07225 and by research funds of
the University of Utah.
\vfill\eject
\taghead{A.}
\line{\bf Appendix:  Boundary conditions for
cylindrically symmetric spacetimes.
\hfill}

In this Appendix we summarize the boundary conditions used in the analysis of
cylindrically symmetric spacetimes.  These boundary conditions guarantee
existence and
differentiability of the various action and Hamiltonian functionals that are
used in the paper, as well as the canonical nature of the various
transformations considered.
The boundary conditions are also such that they are preserved under the
dynamical evolution generated by the Hamiltonian functionals.
It is conceivable that weaker boundary conditions can be used, but
those used here allow for a reasonably large class of solutions to the field
equations.  In particular, the boundary conditions include the Einstein-Rosen
wave solutions which arise when one assumes whole cylindrical symmetry.

It is
important to note that the asymptotic values ($r\rightarrow\infty$) of the
lapse and lapse density, while arbitrary, are
held fixed when varying the Hamiltonian (or action) defined on the ADM phase
space $\Gamma$.  The asymptotic values of the lapse
and lapse density are only allowed to vary when using the extended phase space
$\Gamma^\star$ for cylindrically symmetric spacetimes.  Note also that many
other variables have a non-vanishing value as $r\rightarrow0$ or
$r\rightarrow\infty$; these limiting values are {\it not} held fixed in any of
the variational principles.  In particular, the values of the lapse and lapse
density on the axis $r=0$ are {\it not} held fixed.

The phase space variables are functions of the radial coordinate $r$; we must
give the boundary conditions as
$r\rightarrow0$ and $r\rightarrow\infty$.  As $r\rightarrow0$ we assume the
following behavior of the canonical coordinates and momenta on $\Gamma$:
$$
\eqalignno{
R&=r+ \ord(r^3),&(a1)\cr
\gamma&=\ord(r^2),&(a2)\cr
\psi&=\psi(0)+\ord(r^2),&(a3)\cr
\psit&=\psit(0)+\ord(r^3),&(a4)\cr
\pi_R&=\ord(r),&(a5)\cr
\pi_\gamma&=\ord(r^2),&(a6)\cr
\pi_\psi&=\ord(r),&(a7)\cr
\pi_\psit&=\pi_\psit(0) + \ord(r).&(a8)}
$$
The lapse and shift are assumed to have the following behavior as
$r\rightarrow0$:
$$
\eqalignno{
N^\perp&=N^\perp(0) + \ord(r^2),&(a9)\cr
N^r&=\ord(r^2).&(a10)}
$$
The lapse density inherits its behavior as $r\rightarrow0$ from the lapse
function:
$$
N=N(0) + \ord(r^2).\tag a11
$$
Equations \(a6) and \(253) imply that as
$r\rightarrow0$,
$$
T=T(0)+\ord(r^3).\tag a12
$$
Similarly we assume
$$
\Pi_T={\cal O}(r)\tag a1201
$$
and
$$
\Pi_R={\cal O}(r).\tag a1202
$$
Our boundary conditions at $r=0$ imply that there are no singularities on the
axis of symmetry.

As $r\rightarrow\infty$ we assume the following behavior of the
canonical coordinates and momenta on $\Gamma$:
$$
\eqalignno{
R&=r+ \ord(r^{-\epsilon}),&(a13)\cr
\gamma&=\gamma(\infty)+\ord(r^{-\epsilon}),&(a14)\cr
\psi&=\ord(r^{-\epsilon}),&(a15)\cr
\psit&=\ord(r^{-\epsilon}),&(a16)\cr
\pi_R&=\ord(r^{-1}),&(a17)\cr
\pi_\gamma&=\ord(r^{-(1+\epsilon)}),&(a18)\cr
\pi_\psi&=\ord(r^{-1}),&(a19)\cr
\pi_\psit&=\ord(r^{-(1+\epsilon)}),&(a20)}
$$
where $\epsilon>0$.    The lapse and
shift are assumed to have the following behavior as $r\rightarrow\infty$:
$$
\eqalignno{
N^\perp&=N^\perp(\infty) + \ord(r^{-\epsilon}),&(a22)\cr
N^r&=\ord(r^{-\epsilon}).&(a23)}
$$
The lapse density inherits its behavior as $r\rightarrow\infty$ from the lapse
function:
$$
N=N(\infty) + \ord(r^{-\epsilon}).\tag a24
$$
Equations \(a18) and \(253) imply that as
$r\rightarrow\infty$,
$$
T=T(\infty)+\ord(r^{-\epsilon}).\tag a21
$$
Similarly,
$$
\Pi_T={\cal O}(r^{-(1+\epsilon)})\tag a2101
$$
and
$$
\Pi_R={\cal O}(r^{-1}).\tag a2102
$$
\references

\refis{ADM1962}{R. Arnowitt, S. Deser, and C. Misner in {\it
Gravitation: An
Introduction to Current Research}, edited by L. Witten (Wiley, New York
1962).}

\refis{Ashtekar1994}{A.~Ashtekar and M.~Varadarajan, \prd 50, 4944, 1994.}

\refis{Berger1995}{B.~Berger, P.~Chru\'sciel, and V.~Moncrief, \ann 237,
322, 1995.}

\refis{Dirac1958}{ P.~A.~M.~Dirac, \journal Proc. Roy. Soc., A246, 333, 1958.}

\refis{Dirac1964}{P.A.M. Dirac, {\it Lectures on Quantum Mechanics},
(Yeshiva
University, New York, 1964).}

\refis{Geroch1971a}{R.~Geroch, \jmp 12, 918, 1971.}

\refis{Geroch1971b}{R.~Geroch, \jmp 13, 394, 1971.}

\refis{Gowdy1971}{R.~Gowdy, \prl 27, 826 and 1102, 1971.}

\refis{Henneaux1992}{M. Henneaux and C. Teitelboim, {\it Quantization of Gauge
Systems}, (Princeton University Press, Princeton, 1992).}

\refis{Kuchar1972}{K. V. Kucha\v r, \jmp 13, 758, 1972}.

\refis{Kuchar1992}{K. V. Kucha\v r, ``Time and Interpretations of
Quantum Gravity'', in the proceedings of {\it The Fourth Canadian Conference
on General Relativity and Relativistic Astrophysics}, edited by G. Kunstatter,
D. Vincent, and J. Williams (World Scientific, Singapore 1992).}

\refis{Kuchar1971}{K. V. Kucha\v r, \prd 4, 955,
1971.}

\refis{Kuchar1989b}{K.~Kucha\v r, \prd 39, 2263, 1989.}

\refis{Kuchar1994}{K.~V.~Kucha\v r, \prd 50, 3961, 1994.}

\refis{Moncrief1980}{V.~Moncrief, \ann 132, 87, 1980.}

\refis{Neville1993}{D.~E.~Neville, \cqg 10, 2223, 1993.}

\refis{Teitelboim1984}{C.~Teitelboim, in {\it Quantum Theory of Gravity},
edited by S. Christensen, (Adam Hilger, Bristol, 1984).}

\refis{CGT1989}{K. V. Kucha\v r, C. G. Torre, \jmp 30, 1769, 1989.}

\refis{CGT1991b}{K. V. Kucha\v r and C. G. Torre in {\it Conceptual
Problems
of Quantum Gravity}, edited by A. Ashtekar and J. Stachel, (Birkh\"auser,
Boston 1991).}

\refis{CGT1989a}{C.~G.~Torre, \prd 40, 2588, 1989.}

\refis{CGT1992a}{C. G. Torre, \prd 46, R3231, 1992.}

\refis{CGT1991}{K. V. Kucha\v r and C. G. Torre, \prd 43, 419, 1991.}

\refis{Regge1974}{T.~Regge and C.~Teitelboim, \ann 88, 286, 1974.}

\refis{Gowdy1974}{R.~Gowdy, \ann 83, 203, 1974.}

\refis{York1980}{Y. Choquet-Bruhat and J. York in {\it General
Relativity
and Gravitation: 100 Years After the Birth of Albert Einstein, Vol. 1},
edited by A. Held (Plenum, NY 1980).}

\refis{liealgebranote}{This Lie algebra is isomorphic to the Lie algebra of the
two-dimensional conformal group.  See \refto{CGT1989} for an interpretation of
this result.}

\refis{ctnote}{It is
possible to translate the transformation \(336)--\(3391) into a form applicable
to the
cylindrically symmetric theory, but in this case the boundary conditions
needed on the harmonic maps are somewhat unnatural.}

\endreferences
\bye